\documentclass[letterpaper,twocolumn,10pt]{article}
\usepackage{usenix2019_v3}

\usepackage[utf8]{inputenc}
\usepackage{booktabs} 
\usepackage{graphicx}
\usepackage{subfigure}
\usepackage{units}
\usepackage{blindtext}
\usepackage{xcolor,xspace}
\usepackage{amsmath}
\usepackage{amssymb}
\usepackage{paralist}
\usepackage{amsmath}
\usepackage{tikz}
\usepackage{booktabs}
\usepackage{pifont}
\usepackage{multirow}
\usepackage{hhline}
\usepackage{balance}
\usepackage{footnote}
\usepackage{logicproof}
\usepackage{authblk}
\usepackage{amsthm}
\usetikzlibrary{calc}
\usetikzlibrary{decorations.pathreplacing}

\usepackage{arydshln}
\usepackage{mdframed}
\usepackage{pgf-umlsd}
 \usetikzlibrary{patterns}

\usepackage[T1]{fontenc}
\usepackage{
  beramono,
  listings,
  textcomp
}

\usetikzlibrary{arrows,automata}
\usetikzlibrary{calc}
\usetikzlibrary{arrows,calc,shapes,decorations.pathreplacing}

\usepackage{listings}
\usepackage[ruled]{algorithm2e}
\usepackage{algorithmic}

\LinesNumbered

\usepackage{url}

\usepackage[
    n,
    advantage,
    operators,
    sets,
    adversary,
    landau,
    probability,
    notions,
    logic,
    ff,
    mm,
    primitives,
    events,
    complexity,
    asymptotics,
    keys]{cryptocode}

\newcommand{\newstuff}[1]{#1}

\newcommand{\styleAlgorithm}[1] {\textbf{\texttt{#1}}\xspace}

\newcommand{\OParan}{\styleAlgorithm{(}}
\newcommand{\CParan}{\styleAlgorithm{)}}

\newcommand{\acron}{{{\sf\it VAPE}}\xspace} 
\newcommand{\vrased}{{{\sf\it VRASED}}\xspace}

\newcommand{\dev}{{\ensuremath{\sf{\mathcal Prv}}}\xspace}
\newcommand{\vrf}{{\ensuremath{\sf{\mathcal Vrf}}}\xspace}

\newcommand{\RA}{{\ensuremath{\sf{RA}}}\xspace}

\newcommand{\chal}{{\ensuremath{\sf{\mathcal Chal}}}\xspace}
\newcommand{\xout}{{\ensuremath{\sf{\mathcal O}}}\xspace}

\newcommand{\pox}{\texttt{PoX}\xspace}
\newcommand{\request}{\ensuremath{\mathsf{XRequest}}\xspace}
\newcommand{\execute}{\ensuremath{\mathsf{XAtomicExec}}\xspace}
\newcommand{\prove}{\ensuremath{\mathsf{XProve}}\xspace}
\newcommand{\vrfy}{\ensuremath{\mathsf{XVerify}}\xspace}
\newcommand{\Xrequest}{\ensuremath{\mathsf{XRequest}}\xspace}
\newcommand{\Xexecute}{\ensuremath{\mathsf{XAtomicExec}}\xspace}
\newcommand{\Xprove}{\ensuremath{\mathsf{XProve}}\xspace}
\newcommand{\Xvrfy}{\ensuremath{\mathsf{XVerify}}\xspace}

\newcommand{\xsw}{{\ensuremath{\sf{\mathcal S}}}\xspace}
\newcommand{\token}{{\ensuremath{\sf{\mathcal H}}}\xspace}

\newcommand{\attkey}{\ensuremath{\mathcal K}\xspace}

\renewcommand\adv{\ensuremath{\sf{\mathcal Adv}}\xspace}

\mathchardef\mhyphen="2D

\newcommand{\hw}{\texttt{\small HW-Mod}\xspace}
\newcommand{\sw}{\texttt{\small SW-Att}\xspace}

\newcommand{\swflat}{\texttt{SW-Att}\xspace}
\newcommand{\rom}{\texttt{ROM}\xspace}
\newcommand{\ram}{\texttt{RAM}\xspace}
\newcommand{\hmac}{HMAC\xspace}

\newcommand{\dmaaddr}{\ensuremath{DMA_{addr}}\xspace}
\newcommand{\dmaen}{\ensuremath{DMA_{en}}\xspace}

\newcommand{\ignore}[1]{}

\newtheorem{definition}{Definition}
\newtheorem{theorem}{Theorem}

\author[1]{Ivan De Oliveira Nunes}
\author[2]{Karim Eldefrawy}
\author[1]{Norrathep Rattanavipanon}
\author[1]{Gene Tsudik}
\affil[1]{University of California, Irvine}
\affil[2]{SRI International}

\affil[ ]{\{ivanoliv, nrattana, gene.tsudik\}@uci.edu, karim.eldefrawy@sri.com}

\begin{document}

\title{A Verified Architecture for Proofs of Execution on
	Remote Devices under Full Software Compromise}

\maketitle

\begin{abstract}
Modern society is increasingly surrounded by, and  is growing accustomed to, a wide range of Cyber-Physical Systems (CPS), 
Internet-of-Things (IoT), and {\em smart} devices. They often perform safety-critical functions, e.g., personal medical devices,
automotive CPS as well as industrial and residential automation (e.g., sensor-alarm combinations). On the lower end of the scale, 
these devices are small, cheap and specialized sensors and/or actuators. 
They tend to be equipped with small anemic CPU, have small amounts of  memory and run simple software.
If such devices are left unprotected, consequences of 
forged sensor readings or ignored actuation commands can be catastrophic, particularly, in safety-critical settings.
This prompts the following three questions:  
\textit{(1) How to trust data produced, or verify that commads were performed, by a simple remote embedded device?},
\textit{(2) How to bind these actions/results to the execution of expected software?} and, 
\textit{(3) Can (1) and (2) be attained even if all software on a device could be modified and/or compromised?}

In this paper we answer these questions by designing, showing security of, and formally verifying, \acron: 
\underline{V}erified \underline{A}rchitecture for \underline{P}roofs of \underline{E}xecution.
To the best of our knowledge, this is the first of its kind result for low-end embedded systems. Our work has a range of applications, 
especially, to authenticated sensing and trustworthy actuation, which are increasingly relevant in the context of safety-critical systems.
\acron architecture is publicly available and our evaluation indicates that it incurs low overhead, affordable even 
for very low-end embedded devices, e.g., those based on TI MSP430 or AVR ATmega processors.
\end{abstract}

\section{Introduction}\label{intro}
The number and diversity of special-purpose computing devices has been increasing dramatically. This includes 
all kinds of embedded devices, cyber-physical systems (CPS) and Internet-of-Things (IoT) gadgets, utilized in various 
``smart'' or instrumented settings, such as homes, offices, factories, automotive systems and public venues. Tasks performed by 
these devices are often safety-critical. For example, a typical industrial control system depends on physical measurements 
(e.g., temperature, pressure, humidity, speed) reported by sensors, and on actions taken by actuators, 
such  as: turning on the A/C, sounding an alarm, or reducing speed.

A cyber-physical control system is usually composed of multiple sensors and actuators, at the core of each is a low-cost 
micro-controller unit (MCU). Such devices typically run simple software,  often on "bare metal", i.e., with no microkernel or hypervisor. 
They tend to be operated by a remote central control unit. Despite their potential importance to overall system functionality, low-end devices
are typically designed to minimize cost, physical size and energy consumption, e.g., TI MSP430.

Therefore, their architectural security is usually primitive or non-existent, thus making them vulnerable to malware infestations 
and other malicious software modifications.  A compromised MCU can spoof sensed quantities or ignore actuation 
commands, leading to potentially catastrophic results. For example, in a {\em smart} city, large-scale erroneous reports of 
electricity consumption by smart meters might lead to power outages. A medical device that returns incorrect values when 
queried by a remote physician might result in a wrong drug being prescribed to a patient. A compromised car engine 
temperature sensor that reports incorrect (low) readings can lead to undetected overheating and major damage.
However, despite very real risks of remote software compromise, most users believe that these devices execute expected 
software and thus perform their expected function. 

In this paper, we argue that \textbf{Proofs of Execution (\pox)} are both important and necessary for securing low-end MCUs.
Specifically, we demonstrate in Section~\ref{sec:auth_sensing}, that \pox schemes can be used to construct sensors and actuators 
that ``cannot lie'', even under the assumption of full software compromise. In a nutshell, a \pox conveys that an  untrusted remote (and possibly
compromised) device really executed specific software, and all execution results are authenticated and cryptographically bound to 
this execution. This functionality is similar to authenticated outputs that can be produced by software execution in SGX-alike 
architectures~\cite{sgx, sanctum}. However, such architectures are comparatively heavy-weight and unsuitable for low-end devices.

One key building block in designing \pox schemes is Remote Attestation (\RA). Basically, \RA is a means to detect malware on
a remote low-end MCU. It allows a trusted verifier (\vrf) to remotely measure memory contents (or software state) of an untrusted 
embedded device (\dev). \RA is usually realized as a 2-message challenge-response protocol:
\begin{compactenum}
	\item \vrf sends an attestation request containing a challenge (\chal) to \dev. It might also contain a token derived from a 
		secret (shared by \vrf and \dev) that allows \dev to authenticate \vrf.
	\item \dev receives the attestation request, authenticates the token (if present) and computes an {\em authenticated integrity check} 
		over its memory and \chal. The memory region can be either pre-defined, or explicitly specified in the request.
	\item \dev returns the result to \vrf. 
	\item \vrf receives the result, and decides whether it corresponds to a valid memory state.
\end{compactenum}
The {\em authenticated integrity check} is typically realized as a Message Authentication Code (MAC) computed 
over \dev memory. We discuss one concrete \RA architecture in Section~\ref{sec:preliminaries}.

Despite major progress and several proposals for \RA architectures with different assumptions and guarantees~\cite{seshadri2006scuba,perito2010secure,Viper2011,smart,trustlite,tytan,hydra,brasser2016remote,
Sancus17,erasmus,SAP,smarm,ibrahim2017seed,ConAsiaCCS18,vrasedp}, \textbf{\RA alone is insufficient to obtain proofs of execution}. 
\RA allows \vrf to ascertain integrity of software residing in \dev attested memory region. However, \RA by itself offers no guarantee that 
the attested software is ever executed or 
that any such execution completes successfully. Even if the attested software is executed, there is no guarantee that it has 
not been modified (e.g., by malware residing elsewhere in memory) in the time between its execution and its attestation.
This phenomenon is well known as the Time-Of-Check-Time-Of-Use (TOCTOU) problem.
Finally, \RA does not guarantee authenticity and integrity of any output produced by the execution of the attested software.

To bridge this gap, we design and implement \acron: \underline{V}erified \underline{A}rchitecture for \underline{P}roofs of 
\underline{E}xecution.  In addition to \RA, \acron allows \vrf to request an unforgeable proof that the attested software executed 
successfully and (optionally) produced certain authenticated output. These guarantees hold even in case of full software 
compromise on \dev. Our intended contributions are:\\
\noindent -- \textbf{New security service:}
we design and implement \acron for unforgeable remote proofs of execution (\pox). 
\acron is composed with \vrased~\cite{vrasedp}, a formally verified hybrid \RA architecture.
As we discuss in the rest of this paper, obtaining provably secure \pox requires significant architectural support in addition to a secure \RA functionality (see Section~\ref{sec:eval});
Nonetheless, we show that \acron careful design achieves all necessary properties for secure \pox at fairly low overhead.
To the best of our knowledge, this is the first security architecture for proofs of remote software execution on low-end devices.

\noindent-- \textbf{Provable security \& implementation verification:}
secure \pox involves reasoning about several details which can be easily overlooked.
Ensuring that all necessary \pox components are correctly implemented, composed, and integrated with the underlying \RA 
functionality is not trivial. In particular, early \RA architectures oversimplified \pox requirements, leading to the incorrect 
conclusion that \pox can be obtained directly from \RA; see Section~\ref{sec:RW} for examples.
In this work, we prove that \acron yields a secure \pox architecture. All security properties expected from \acron implementation 
are formally specified using Linear Temporal Logic (LTL) and \acron modules are verified to adhere to these properties. 
We also prove that the composition of \acron new modules with a formally verified \RA architecture (\vrased) implies a concrete 
definition of \pox security.

\noindent-- \textbf{Evaluation, publicly available implementation and applications:} \acron was implemented on a real-world low-end 
MCU (TI MSP430) and deployed using commodity FPGAs. Both design and verification are publicly available at \cite{public-code}.
Our evaluation shows low hardware overhead, affordable even for low-end MCUs. The implementation is 
accompanied by a sample \pox application; see Section~\ref{sec:auth_sensing}.
As a proof of concept, we use \acron to construct a trustworthy safety-critical device, on which malware can not 
spoof execution results (e.g., spoof sensed values) without detection.\\

\noindent \textbf{Targeted Devices \& Scope:}
This work focuses on CPS/IoT sensors and actuators with relatively low computing power.
These are some of the lowest-end devices based on low-power single core MCUs with only a few KBytes 
of program and data memory. Two prominent examples are: TI MSP430 and Atmel AVR ATmega.
These are $8$- and $16$-bit CPUs, typically running at $1$-$16$MHz clock frequencies, with $\approx64$ KBytes of addressable memory.
SRAM is used as data memory and its size is normally within $4$-$16$KBytes with the rest of address space available for program memory. 
These devices execute instructions in place (in physical memory) and have no memory management unit (MMU) to support virtual memory.
Our implementation focuses on MSP430. This choice is due to public availability of a well-maintained open-source MSP430 
hardware design from Open Cores \cite{openmsp430}. Nevertheless, our machine model and the entire methodology developed in 
this paper are applicable to other low-end MCUs in the same class, such as Atmel AVR ATmega.
\\

\noindent \textbf{Organization:}
Section~\ref{sec:RW} discusses related work on remote attestation, formal verification of security services and 
control flow attestation. Section \ref{sec:preliminaries} provides some background on automated verification, 
and \vrased \RA architecture. Section \ref{sec:pox_def} introduces Proofs of Execution (\pox), followed by a 
realization thereof in Section \ref{sec:overview}, including technical details of \acron design, as well as the 
adversarial model and assumptions. Section \ref{sec:verif} presents \acron's formal verification.
Next, in Section \ref{sec:eval}, we report \acron's evaluation results 
and describe how to use \acron to implement authenticated sensing/actuation.
Section~\ref{sec:conclusion} concludes the paper with a summary of results.

\section{Related Work}\label{sec:RW}
\noindent \textbf{Remote Attestation (\RA)--} architectures fall into three categories: hardware-based, 
software-based, or hybrid. Hardware-based~\cite{PFM+04, TCG, KKW+12} relies on dedicated secure
hardware components, e.g., Trusted Platform Modules (TPMs)~\cite{tpm}. However, the cost of such hardware is normally prohibitive for 
low-end IoT/CPS devices. Software-based attestation~\cite{KeJa03, SPD+04, SLS+05} requires no hardware
security features but imposes strong security assumptions about communication between \dev and \vrf, which are 
unrealistic in the IoT/CPS ecosystem (though, it is the only choice for legacy devices).  Hybrid 
\RA~\cite{hydra, ETF+12, KSS+14, FNR+14, tytan} aims to achieve security equivalent to hardware-based mechanisms  
at minimal cost. It thus entails minimal hardware requirements while relying on software to reduce overall complexity
and \RA footprint on \dev.

The first hybrid \RA architecture -- SMART~\cite{smart} -- acknowledged the importance of proving remote code execution on \dev, 
in addition to just attesting \dev's memory. Using an \emph{attest-then-execute} approach (see Algorithm 4 in \cite{smart}), 
SMART attempts to achieve software execution 
guarantees by specifying the address of the first instruction to be executed after completion of attestation.
We consider this to be a best-effort approach which merely guarantees that the code {\bf starts executing}.
However, it \emph{does not guarantee that execution completes successfully}.
For example, SMART's approach cannot detect if execution is interrupted (e.g., by malware) and never resumed.
It also cannot detect when a reset (e.g., due to software bugs, or \dev running low on power) happens during 
execution, thus preventing its successful completion. Furthermore, direct memory access (DMA) can occur during 
execution and it can modify the code being executed, or its output. In other words, SMART offers no 
guarantees beyond ``invoking the executable''.

\newstuff{Another notable \RA architecture is TrustLite~\cite{trustlite}, which builds upon SMART to allow secure interrupts.
However, TrustLite does not enforce temporal consistency of attested memory;  it is 
thus conceptually vulnerable to self-relocating malware and  memory modification during attestation~\cite{carpent2018temporal}. 
Consequently, it is challenging to deriving secure \pox from TrustLite. Several other prominent low-to-medium-end \RA
architectures -- e.g., SANCUS~\cite{Sancus17}, HYDRA~\cite{hydra}, and TyTaN~\cite{tytan} -- do not offer \pox.
In this paper, we show that the \emph{execute-then-attest} approach, using a temporally consistent \RA architecture,
provides unforgeable proofs of execution that are produced only if the execution and its results are  not tampered with, 
and it completes successfully.}

\noindent \textbf{Control Flow Attestation (CFA)--} In contrast with \RA, which measures \dev's software integrity, CFA techniques~\cite{cflat,dessouky2017fat,zeitouni2017atrium,dessouky2018litehax} provide \vrf with a measurement of 
the exact control flow path taken during execution of specific software on \dev. Such measurements allow \vrf 
to detect run-time attacks. We believe that it is possible to construct a \pox scheme that relies on CFA to produce 
proofs of execution based on the attested control flow path. However,  in this paper, 
we advocate a different approach -- specific for proofs of execution -- for two main reasons:
\begin{compactitem}
 \item CFA requires substantial additional hardware features in order to attest, in real time, executed instructions 
 along with memory addresses and the program counter. For example, C-FLAT~\cite{cflat} assumes ARM TrustZone, 
 while LO-FAT~\cite{dessouky2017fat} and LiteHAX\cite{dessouky2018litehax} require a branch monitor and a hash 
 engine. We believe that such hardware components are not viable for low-end devices,
 since their cost (in terms of price, size, and energy consumption) is typically higher than the cost of a low-end MCU 
 itself. For example, the cheapest Trusted Platform Module (TPM)~\cite{tpm}, is about $10\times$ more expensive 
 than MSP430 MCU itself\footnote{Source: https://www.digikey.com/}. As shown in Section~\ref{sec:experiments}, current 
 CFA architectures are also considerably more expensive than the MCU itself and hence not realistic in our device context.
 \item CFA assumes that \vrf can enumerate a large (potentially exponential!) number of valid control flow paths for a given 
 program, and verify a valid response for each. This burden is unnecessary for determining if a proof of execution is valid, 
 because one does not need to know the exact execution path in order to determine if execution occurred (and terminated) 
 successfully (see Section~\ref{sec:adv_model} for a discussion on run-time threats).
\end{compactitem}
Instead of relying on CFA, our work constructs a \pox-specific architecture -- \acron -- that enables low-cost \pox for 
low-end devices. \acron is non-invasive (i.e.,it does not modify MCU behavior and semantics) and incurs low hardware 
overhead: around 2\% for registers and 12\% for LUTs. Also,  \vrf is not required to enumerate valid control flow graphs 
and the verification burden for \pox is exactly the same as the effort to verify a typical remote attestation response 
for the same code.

\noindent \textbf{Formally Verified Security Services--} 
In recent years, several efforts focused on formally verifying security-critical systems. In terms of cryptographic primitives, 
Hawblitzel et al. \cite{hawblitzel2014ironclad} verified implementations of SHA, HMAC, and RSA. 
Bond et al.~\cite{bond2017vale} verified an assembly implementation of SHA-256, Poly1305, AES and ECDSA. 
Zinzindohou{\'e}, et al.~\cite{hacl} developed HACL*, a verified cryptographic  library containing the entire cryptographic 
API of NaCl~\cite{bernstein2012security}.  Larger security-critical systems have also been successfully verified. 
Bhargavan~\cite{bhargavan2013implementing} implemented the TLS protocol with verified cryptographic security.
CompCert\cite{compcert} is a C compiler that is formally verified to preserve C code semantics in generated assembly code. 
Klein et al.~\cite{sel4} designed and proved functional correctness of the seL4 microkernel. More recently, 
\vrased~\cite{vrasedp} realized a formally verified hybrid \RA architecture. \acron architecture, proposed in this paper, 
uses \vrased~\RA functionality (see Section~\ref{sec:vrased} for details)  composed with additional formally verified 
architectural components to obtain provably secure \pox.

\section{Background}\label{sec:preliminaries}
\subsection{Formal Verification, Model Checking \& Linear Temporal Logic}\label{sec:prelim-fv}
Computer-aided formal verification typically involves three basic steps. First, the system of interest (e.g., hardware, software, 
communication protocol) is described using a formal model,  e.g., a Finite State Machine (FSM). Second, properties that 
the model should satisfy are formally specified. Third, the system model is checked against formally specified properties to 
guarantee that the system retains them. This can be achieved by either Theorem Proving or Model Checking.
In this work, we use the latter to verify the implementation of system modules, and the former to derive new properties 
from sub-properties that were proved for the modules' implementation.

In one instantiation of model checking, properties are specified as \textit{formulae} using Temporal Logic (TL) and system models 
are represented as FSMs. Hence, a system is represented by a triple $(S, S_0, T)$, where $S$ is a finite set of states,
$S_0 \subseteq S$ is the set of possible initial states, and $T \subseteq S \times S$ is the transition relation set -- it describes 
the set of states that can be reached in a single step from each state. The use of TL to specify properties allows representation 
of expected system behavior over time.

We apply the model checker NuSMV~\cite{cimatti2002nusmv}, which can be used to verify generic HW or SW models.
For digital hardware described at Register Transfer Level (RTL) -- which is the case in this work -- conversion
from Hardware Description Language (HDL) to NuSMV model specification is simple. Furthermore, it can be 
automated~\cite{irfan2016verilog2smv}, because the standard RTL design already relies on describing hardware as an FSM.

In NuSMV, properties are specified in Linear Temporal Logic (LTL), which is particularly useful for verifying sequential 
systems, since LTL extends common logic statements with temporal clauses. In addition to propositional 
connectives, such as conjunction ($\land$), disjunction ($\lor$), negation ($\neg$), and implication ($\rightarrow$), 
LTL includes temporal connectives, thus enabling sequential reasoning. In this paper, we are interested in the following 
temporal connectives:
\begin{compactitem}
 \item \textbf{X}$\phi$ -- ne\underline{X}t $\phi$: holds if $\phi$ is true at the next system state.
 \item \textbf{F}$\phi$ -- \underline{F}uture $\phi$: holds if there exists a future state where $\phi$ is true.
 \item \textbf{G}$\phi$ -- \underline{G}lobally $\phi$: holds if for all future states $\phi$ is true.
 \item $\phi$ \textbf{U} $\psi$ -- $\phi$ \underline{U}ntil $\psi$: holds if there is a future state where $\psi$ holds and
 	$\phi$ holds for all states prior to that.
 \item $\phi$ \textbf{B} $\psi$ -- $\phi$ \underline{B}efore $\psi$: holds if the existence of state where $\psi$ 
 	holds implies the existence of an earlier state where $\phi$ holds. This connective can be expressed using 
	\textbf{U} through the equivalence: $\phi$ \textbf{B} $\psi \equiv \neg(\neg\phi$ \textbf{U} $\psi)$.
\end{compactitem}
This set of temporal connectives combined with propositional connectives (with their usual meanings) allows us to specify 
powerful rules. NuSMV works by checking LTL specifications against the system FSM for all reachable states in such FSM.

\subsection{Formally Verified \RA}\label{sec:vrased}
\vrased~\cite{vrasedp} is a formally verified hybrid (hardware/software co-design) RA architecture, 
built as a set of sub-modules, each guaranteeing a specific set of sub-properties.
All \vrased sub-modules, both hardware and software, are individually verified. Finally, the composition of all sub-modules 
is proved to satisfy formal definitions of \RA soundness and security. {\bf \RA soundness} guarantees that an integrity-ensuring 
function (\hmac in \vrased's case) is correctly computed on the exact memory being attested. Moreover, it guarantees that attested 
memory remains unmodified after the start of \RA computation, protecting against ``hide-and-seek'' attacks caused by self-relocating 
malware~\cite{carpent2018temporal}. {\bf \RA security} ensures that \RA execution generates an unforgeable 
authenticated memory measurement and that the secret key \attkey used in computing this measurement is not leaked before, 
during, or after, attestation.

To achieve aforementioned goals, \vrased software (\sw) is stored in Read-Only Memory (\rom) and relies on a (previously) 
formally verified \hmac implementation from HACL* cryptographic library~\cite{hacl}. A typical execution of \sw is carried out as follows:
\begin{compactenum}
	\item Read challenge \chal from memory region $MR$. 
	\item Derive a one-time key from \chal and the attestation master key \attkey.
	\item Generate an attestation token $H$ by computing an HMAC over an attested memory region $AR$ using the derived key:\\
	\centerline{$H = HMAC(KDF(\attkey, MR), AR)$}
	\item Write $H$ into $MR$ and return the execution to unprivileged software, i.e, normal applications.
\end{compactenum}
\noindent \vrased hardware (\hw) monitors $7$ MCU signals:
\begin{compactitem}
	\item $PC$: Current Program Counter value;
	\item $R_{en}$: Signal that indicates if the MCU is reading from memory (1-bit);
	\item $W_{en}$: Signal that indicates if the MCU is writing to memory (1-bit);
	\item $D_{addr}$: Address for an MCU memory access;
	\item \dmaen: Signal that indicates if  Direct Memory Access (DMA) is currently enabled (1-bit);
	\item \dmaaddr: Memory address being accessed by DMA.
	\item $irq$: Signal that indicates if an interrupt is happening (1-bit);
\end{compactitem}
These signals are used to determine a one-bit $reset$ signal output. Whenever $reset$ is set to $1$ a system-wide 
MCU reset is triggered immediately, i.e., before the execution of the next instruction. This condition is triggered whenever
\vrased's hardware detects any violation of its security properties. \vrased hardware is described in Register Transfer Level 
(RTL) using Finite State Machines (FSMs). Then, NuSMV Model Checker~\cite{nusmv} is used to automatically prove that 
such FSMs achieve claimed security sub-properties. Finally, the proof that the conjunction of hardware and software sub-properties 
implies end-to-end soundness and security is done using an LTL theorem prover. More formally, \vrased end-to-end security 
proof guarantees that no probabilistic polynomial time (PPT) adversary can win the \RA \emph{security game}
with non-negligible probability in terms of the security parameter. (See Definition~\ref{def:vrased_sec} in Appendix~\ref{apdx:prove}).

\begin{figure*}[!ht]
\begin{mdframed}
\begin{definition}[Proof of Execution (\pox) Scheme]~\\
\label{def:pox}
\footnotesize
A Proof of Execution (\pox) scheme is a tuple of algorithms $[\Xrequest, \Xexecute, \Xprove, \Xvrfy]$ 
performed between \dev and \vrf where:\\
\begin{compactenum}
%
\item{$\Xrequest^{\vrf \rightarrow \dev}(\xsw, \cdot)$:} is an algorithm executed by \vrf which takes as 
\underline{input} some software \xsw (consisting of a list of instructions $\{s_1, s_2, ..., s_m\}$).
\vrf expects an honest \dev to execute \xsw.
$\request$  generates a challenge $\chal$, and embeds it 
alongside \xsw, into an \underline{output} request message asking \dev to execute \xsw, and to prove that such 
execution took place.\\
%
%
\item{$\Xexecute^{\dev}(ER, \cdot)$:} an algorithm (with possible hardware-support) that takes as \underline{input} some executable region $ER$ in \dev's memory,
 containing a list of instructions $\{i_1, i_2, ..., i_m\}$.
\Xexecute runs on \dev and is considered successful iff: (1) instructions in $ER$ are executed from its first 
instruction, $i_1$, and end at its last instruction, $i_m$; (2) ER's execution is atomic, i.e.,
if $E$ is the sequence of instructions executed between $i_1$ and $i_m$, then $\{ e | e \in E\} \subseteq ER$;
and (3) $ER$'s execution flow is not altered by external events, i.e., MCU interrupts or DMA events.
The $\Xexecute$ algorithm \underline{outputs} a string \xout. Note that  \xout may be a default string ($\perp$) if $ER$'s execution 
does not result in any output.\\
\item{$\Xprove^{\dev}(ER, \chal, \xout, \cdot)$:} an algorithm (with possible hardware-support) that takes as \underline{input} 
some $ER$, \chal and \xout and is run by \dev to \underline{output} \token, i.e., a proof that 
$\Xrequest^{\vrf \rightarrow \dev}(\xsw, \cdot)$ and $\Xexecute^{\dev}(ER, \cdot)$ happened (in this sequence) 
and that \xout was produced by $\Xexecute^{\dev}(ER, \cdot)$.\\
\item{$\Xvrfy^{\dev \rightarrow \vrf}(\token, \xout, \xsw, \chal, \cdot)$:} an algorithm executed by $\vrf$ with the 
following \underline{inputs}: some \xsw, \chal, \token and \xout. 
The \Xvrfy algorithm checks whether $\token$ is a valid proof of the execution of \xsw (i.e., executed memory region $ER$ corresponds to \xsw) on \dev given the challenge \chal, 
and if \xout is an authentic output/result of such an execution. If both checks succeed, \Xvrfy \underline{outputs} \texttt{1}, 
otherwise it outputs \texttt{0}.\\

\noindent\textbf{Remark:} In the parameters list, ($\cdot$) denotes that additional parameters might be included depending on the specific \pox construction.
\end{compactenum}
\end{definition}
\end{mdframed}
\normalsize
\end{figure*}

\begin{figure*}[!ht]
\begin{mdframed}
\begin{definition}[\pox Security Game]\label{def:pox_sec_def}~\\
\footnotesize
-- Let $t_{req}$ denote time when \vrf issues $\chal \leftarrow \Xrequest^{\vrf \rightarrow \dev} (\xsw)$.\\
-- Let $t_{verif}$ denote time when \vrf receives \token and \xout back from \dev in response to $\Xrequest^{\vrf \rightarrow \dev}$.\\
-- Let $\Xexecute^{\dev}(\xsw,t_{req} \rightarrow t_{verif})$ denote that $\Xexecute^{\dev}(ER, \cdot)$, such that $ER \equiv \xsw$, was invoked and 
	completed within the time interval $[t_{req},t_{verif}]$. \\
-- Let $\xout \equiv \Xexecute^{\dev}(\xsw,t_{req} \rightarrow t_{verif})$ denote that $\Xexecute^{\dev}(\xsw,t_{req} 
	\rightarrow t_{verif})$ produces output $\xout$.
Conversely, $\xout \not\equiv \Xexecute^{\dev}(\xsw,t_{req} \rightarrow t_{verif})$ indicates $\xout$ is not produced by 
$\Xexecute^{\dev}(\xsw,t_{req} \rightarrow t_{verif})$. \\
\textbf{\ref{def:pox_sec_def}.1 \pox Security Game (\pox-game):}
Challenger plays the following game with \adv:
	\begin{compactenum}
	\item \adv is given full control over \dev software state 
	and oracle access to calls to the algorithms $\Xexecute^{\dev}$ and $\Xprove^{\dev}$.
	\item At time $t_{req}$, \adv is presented with software \xsw and challenge $\chal$.
	\item \adv wins in two cases:
	  \begin{compactenum}
	    \item \textbf{None or incomplete execution:} \adv produces ($\token_{\adv}$,$\xout_{\adv}$), 
	    	such that $\vrfy(\token_{\adv}, \xout_{\adv}, \xsw, \chal, \cdot) = 1$,\\ without calling $\execute^{\dev}(\xsw,t_{req} \rightarrow t_{verif})$.
	    \item \textbf{Execution with tampered output:} \adv calls $\execute^{\dev}(\xsw,t_{req} \rightarrow t_{verif})$ and can produce 
	    	($\token_{\adv}$,$\xout_{\adv}$),\\ such that $\vrfy(\token_{\adv}, \xout_{\adv}, \xsw, \chal, \cdot) = 1$ and  
		$\xout_{\adv} \not\equiv \execute^{\dev}(\xsw,t_{req} \rightarrow t_{verif})$
	  \end{compactenum}
	\end{compactenum}~
	
\textbf{\ref{def:pox_sec_def}.2 \pox Security Definition:}\\
A \pox scheme is considered secure for security parameter $l$ if, for all PPT adversaries \adv, there exists a negligible function $\negl[]$ such that:
	\begin{center}
		$Pr[\adv, \text{\pox-game}] \leq \negl[l]$
	\end{center}
\end{definition}
\end{mdframed}
\end{figure*}

\section{Proof of Execution ($\pox$) Schemes}\label{sec:pox_def}
A \emph{Proof of Execution} (\pox) is a scheme 
involving two parties: (1)~a trusted verifier \vrf, and (2)~an untrusted (potentially infected) remote prover \dev.
Informally, the goal of \pox is to allow \vrf to request execution of specific software $\xsw$ by \dev.
As part of \pox, \dev must reply to \vrf with an authenticated unforgeable cryptographic proof (\token) that convinces \vrf
that \dev indeed executed \xsw. To accomplish this, verifying \token must prove that: (1)~\xsw executed atomically, in its entirety, and 
that such execution occurred on \dev (and not on some other device); and (2) any claimed result/output value of such execution, 
that is accepted as legitimate by \vrf, could not have been spoofed or modified. Also, the size and behavior (i.e., instructions) 
of \xsw, as well as the size of its output (if any), should be configurable and optionally specified by \vrf.
In other words, \pox should provide proofs of execution for arbitrary (including possibly buggy) 
software, along with corresponding authenticated outputs. Definition~\ref{def:pox} specifies \pox schemes in more detail.

We now justify the need to include atomic execution of \xsw in the definition of \pox. 
On low-end MCUs, software typically runs on ``bare metal" and, in most cases, there is no mechanism to enforce memory 
isolation between applications. Therefore, allowing \xsw execution to be interrupted would permit other (potentially malicious) 
software running on \dev to alter the behavior of \xsw. This might be done, for example, by an application that interrupts 
execution of \xsw  and changes intermediate computation results in \xsw data memory, thus tampering with its output or control flow.
Another example is an interrupt that resumes \xsw at different instruction modifying \xsw execution flow. Such actions
could modify \xsw behavior completely via return oriented programming (ROP).

\subsection{\pox Adversarial Model \& Security Definition}\label{sec:adv_model}
We consider an adversary \adv that might control \dev's entire software state, code, and data.
\adv\ can modify any writable memory and read any memory that is not explicitly protected by hardware-enforced
access control rules. 
\adv\ may also have full control over all Direct Memory Access (DMA) controllers of \dev. Recall that DMA allows a hardware controller 
to directly access main memory (e.g., \ram, flash or \rom) without going through the CPU.

We consider a scheme $\pox$ $=$ $(\Xrequest,$ $\Xexecute,$ $\Xprove,$ $\Xvrfy)$ to be secure
if the aforementioned \adv has only negligible probability of convincing \vrf that \xsw executed successfully when, in 
reality,  such execution did not take place, or was interrupted. In addition we require that, if execution of 
\xsw occurs, \adv cannot tamper with, or influence, this execution's outputs.
These notions are formalized by the security game in Definition~\ref{def:pox_sec_def}.

We note that Definition~\ref{def:pox_sec_def} binds execution of \xsw to the time between \vrf issuing 
the request and receiving the response. Therefore, if a \pox scheme is secure according to this definition, 
\vrf can be certain about freshness of the execution. In the same vein, the output produced by such execution 
is also guaranteed to be fresh. This timeliness property is important to avoid replays of previous valid 
executions; in fact, it is essential for safety-critical applications. See Section~\ref{sec:auth_sensing} for examples.

\textbf{Correctness of the Executable:} we stress that the purpose of \pox is to offer a guarantee that \xsw, as specified by \vrf, was executed.
Similar to Trusted Execution Environments targeting high-end CPUs, such as Intel SGX, \pox schemes do not aim to 
check correctness and absence of implementation bugs in \xsw. As such, it is not concerned with run-time attacks that 
exploit bugs and vulnerabilities in \xsw implementation itself, to change its expected behavior (e.g., by executing \xsw with 
inputs crafted to exploit \xsw bugs and hijack its control flow). In particular, correctness of \xsw need \textbf{not} be assured 
by the low-end \dev. Since \vrf is a more powerful device and knows \xsw, 
it has the ability (and more computational resources) to employ various vulnerability detection methods
(e.g., fuzzing~\cite{fuzzer} or static analysis~\cite{costin2014large}) or even software formal verification (depending on the level of 
rigor desired) to avoid or detect implementation bugs in \xsw.
This type of techniques can be performed offline before sending \xsw to \dev and the whole issue is orthogonal to \pox functionality.

\textbf{Physical Attacks:} physical and hardware-focused attacks are out of scope of this paper. Specifically, we assume that
\adv\ cannot modify code in \rom, induce hardware faults, or retrieve \dev secrets via physical presence 
side-channels. Protection against such attacks is considered orthogonal and could be supported via standard 
physical security techniques~\cite{ravi2004tamper}.

\subsection{MCU Assumptions}\label{sec:adv}
\label{sec:adv_model}
\acron is composed with \vrased to enable a verified architecture for proofs of execution.
Therefore, we assume the same machine model introduced in \vrased and make no additional assumptions.
We review these assumptions throughout the rest of this section
and then formalize them as an LTL machine model in Section~\ref{sec:verif}.

Verification of the entire CPU is beyond the scope of this paper.
Therefore, we assume the CPU architecture strictly adheres to, and correctly implements, its specifications.
In particular, our design and verification rely on the following simple axioms:\\
\noindent {\bf A1 -- \emph{Program Counter (PC):}} $PC$ always contains the address of the instruction being executed in a given CPU cycle.

\noindent {\bf A2 -- \emph{Memory Address:}} Whenever memory is read or written, a data-address signal ($D_{addr}$) contains 
 the address of the corresponding memory location. For a read access, a data read-enable bit ($R_{en}$) must be set, while, 
 for a write access, a data write-enable bit ($W_{en}$) must be set.
 
\noindent {{\bf A3 -- \emph{DMA:}} Whenever the DMA controller attempts to access the main system memory, a DMA-address signal 
 (\dmaaddr) reflects the address of the memory location being accessed and a DMA-enable bit (\dmaen) must be set. 
 DMA cannot access memory when \dmaen is off (logical zero).
 
\noindent {\bf A4 -- \emph{MCU Reset:}} At the end of a successful reset routine, all registers (including $PC$) are set to zero 
 before resuming normal software execution flow. Resets are handled by the MCU in hardware. Thus, the reset handling routine 
 cannot be modified. When a reset happens, the corresponding $reset$ signal is set. The same signal is also set when 
 the MCU initializes for the first time.
 
\noindent {\bf A5 -- \emph{Interrupts:}} Whenever an interrupt occurs, the corresponding $irq$ signal is set.

\section{\acron: A Secure $\pox$ Architecture}\label{sec:overview}
%

\begin{figure*}[!ht]
\begin{mdframed}
\begin{definition}[Proof of Execution Protocol]\label{def:vape}
\footnotesize
        \acron instantiates a $\pox$ $=$ $(\Xrequest,$ $\Xexecute,$ $\Xprove,$ $\Xvrfy)$ scheme behaving as follows: 
        \begin{compactenum}~
                \item $\Xrequest^{\vrf \rightarrow \dev}\OParan\xsw, ER_{min}, ER_{max}, OR_{min}, OR_{max}\CParan$: includes a 
                set of configuration parameters $ER_{min}$, $ER_{max}$, $OR_{min}$, $OR_{max}$.
                The Executable Range ($ER$) is a contiguous memory block in which \xsw is to be installed: $ER = [ER_{min}, ER_{max}]$.
                Similarly, the Output Range ($OR$) is also configurable and defined by \vrf's request as $OR = [OR_{min}, OR_{max}]$. 
                If \xsw does not produce any output $OR_{min} = OR_{max} = \perp$.
                \xsw is the software to be installed in $ER$ and executed. If $\xsw$ is unspecified ($\xsw = \perp$) the protocol will 
                execute whatever code was pre-installed on $ER$ on \dev, i.e., \vrf is not required to provide \xsw in every request, 
                only when it wants to update $ER$ contents before executing it.
                If  the code for \xsw is sent by \vrf, untrusted auxiliary software in \dev is responsible for copying \xsw into $ER$.
                \dev also receives a random $l$-bit challenge \chal ($|\chal| = l$) as part of the request, where $l$ is the security parameter.\\
                
                \item $\Xexecute^{\dev}\OParan ER, OR, METADATA \CParan$: This algorithm starts with unprivileged auxiliary 
                software writing the values of: $ER_{min}$, $ER_{max}$, $OR_{min}$, $OR_{max}$ and \chal to a special 
                pre-defined memory region denoted by $METADATA$.
                \acron's verified hardware enforces immutability, atomic execution and access control rules according to the values 
                stored in $METADATA$; details are described in Section~\ref{sec:arch}.
                Finally, it begins execution of \xsw by setting the program counter to the value of $ER_{min}$.\\
                \item $\Xprove^{\dev}\OParan ER, \chal, OR \CParan$: produces proof of execution \token. 
                \token allows \vrf to decide whether: (1) code contained in $ER$ actually executed; (2) $ER$ contained specified 
                (expected) \xsw's code during execution; (3) this execution is fresh, i.e., performed after the most recent \request; and 
                (4) claimed output in $OR$ is indeed produced by this execution. As mentioned earlier, \acron uses \vrased's~ \RA architecture 
                to compute \token by attesting at least the executable, along with its output, and corresponding execution metadata. 
                More formally:\\
                \begin{equation}\label{eq:pox_token}
                 \token = \hmac(KDF(\attkey, \chal),ER,OR,METADATA,...)
                \end{equation}\\
                $METADATA$ also contains the $EXEC$ flag that is \textbf{read-only to all software running in \dev and can only be 
                written to  by \acron's formally verified hardware}. This hardware monitors execution and sets $EXEC=1$ only if 
                ER executed successfully (\execute) and memory regions of $METADATA$, $ER$, and $OR$ 
                were not modified between the end of $ER$'s execution and the computation of \token.
                The reasons for these requirements are detailed in Section~\ref{sec:properties}.
                If any malware residing on \dev attempts to violate any of these properties \acron's verified hardware (provably) sets
                $EXEC$ to zero. After computing \token, \dev returns it and contents of $OR$ (\xout) produced by ER's execution to \vrf.  \\
		\item $\Xvrfy^{\dev \rightarrow \vrf}\OParan \token, \xout, \xsw, METADATA_{\vrf}\CParan:$ Upon receiving \token and \xout, 
		\vrf checks whether \token is produced by a legitimate execution of \xsw and reflects parameters specified in \request, i.e., 
		$METADATA_{\vrf} = \chal||OR_{min}||OR_{max}||ER_{min}||ER_{max}|| EXEC=1$.
		This way, \vrf concludes that \xsw successfully executed on \dev and produced output \xout if:
		 \begin{equation}\label{eq:expected_token}
		   \token \equiv \hmac(KDF(\attkey, \chal_{\vrf}), \xsw , \xout , METADATA_{\vrf},...)		  
		 \end{equation}                
       \end{compactenum}
\end{definition}
\end{mdframed}
\end{figure*}

\begin{table}[!htb]
\caption{Notation}
\begin{center}
\scriptsize
\begin{tabular}{r p{6.3cm} }
\toprule
$PC$					&  Current Program Counter value			\\
$R_{en}$					&  Signal that indicates if the MCU is reading from memory (1-bit)		\\
$W_{en}$					&  Signal that indicates if the MCU is writing to memory (1-bit)		\\
$D_{addr}$				&  Address for an MCU memory access 		\\
\dmaen					&  Signal that indicates if DMA is currently enabled (1-bit)				\\
\dmaaddr					&  Memory address being accessed by DMA, if any 				\\
$irq$						&  Signal that indicates if an interrupt is happening 				\\
$CR$					&  Memory region where \swflat is stored: $CR = [CR_{min}, CR_{max}]$   \\
$MR$					& (MAC Region) Memory region in which \swflat computation result is written: $MR = [MR_{min}, MR_{max}]$.
						The same region is used to pass the attestation challenge as input to \swflat \\
$AR$					& (Attested Region) Memory region to be attested.
						Can be fixed/predefined or specified in an authenticated request from \vrf: $AR = [AR_{min}, AR_{max}]$ \\
$KR$					& (Key Region) Memory region that stores \attkey \\
$XS$					& (Exclusive Stack Region) Exclusive memory region that contains \swflat's stack and can be only accessed by \swflat \\
$reset$ 					& A 1-bit signal that reboots/resets the MCU when set to logical $1$ \\
$ER$					& (Execution Region) Memory region that stores an executable to be executed: $ER = [ER_{min}, ER_{max}]$ \\
$OR$					& (Output Region) Memory region that stores execution  output: $OR = [OR_{min}, OR_{max}]$ \\
$EXEC$					& 1-bit execution flag indicating whether a successful execution has happened \\
$METADATA$				& Memory region containing \acron's metadata \\
\bottomrule
\end{tabular}
\end{center}
\label{tab:notation}
\end{table}

We now present \acron, a new \pox architecture that realizes the  \pox security definition in Definition~\ref{def:pox_sec_def}. 
The key aspect of \acron is a computer-aided formally verified and publicly available implementation thereof. This section first 
provides some intuition behind \acron's design. All \acron properties are overviewed informally in this section and are later
formalized in Section~\ref{sec:verif}.

In the rest of this section we use the term ``unprivileged software'' to refer to any software other than \sw code from \vrased.
\adv is allowed to overwrite or bypass any ``unprivileged software''. Meanwhile, ``trusted software'' refers to \vrased's implementation 
of \sw (see Section~\ref{sec:preliminaries}) which is formally verified and can not be modified by \adv, since it is stored in \rom.
\acron is designed such that no changes to \sw are required. Therefore, both functionalities (\RA and \pox, i.e., \vrased and \acron) 
can co-exist on the same device without interfering with each other. 

Notation is summarized in Table~\ref{tab:notation}.

\subsection{Protocol and Architecture}\label{sec:arch}
\begin{figure}[!h]
	\includegraphics[width=\linewidth]{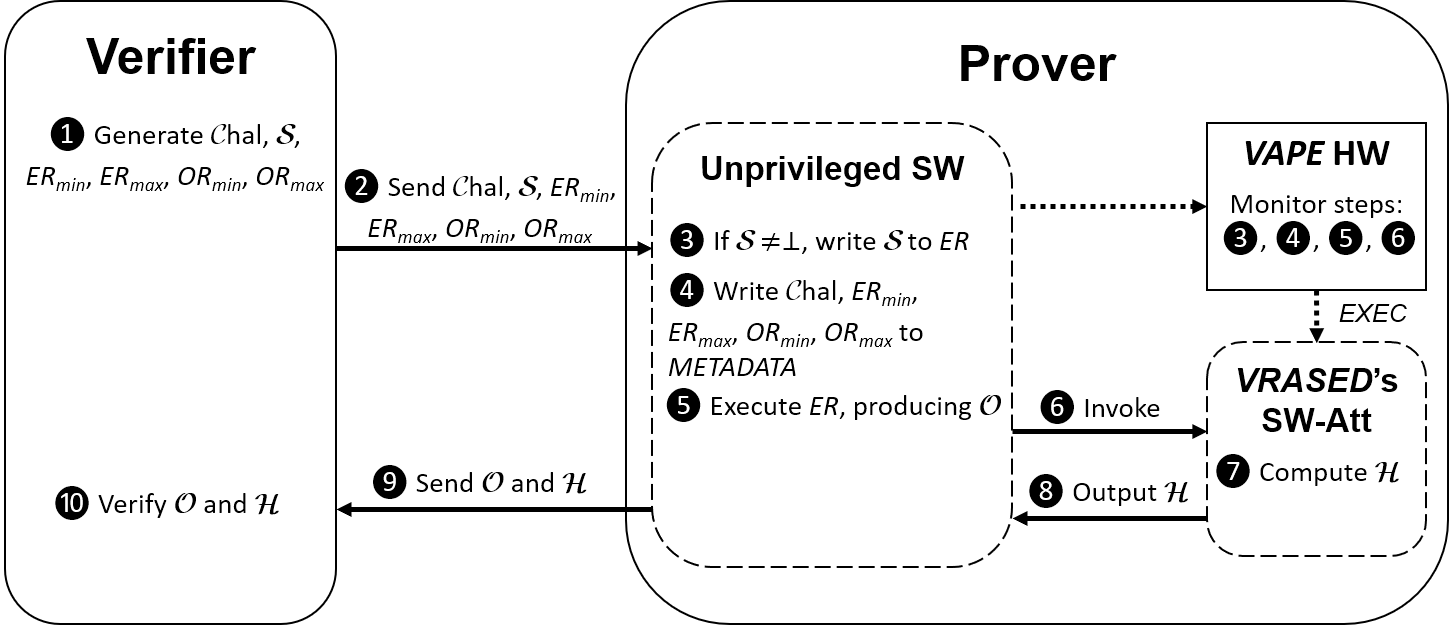}
	\caption{Overview of \acron's workflow}
	\label{fig:seq-events}
\end{figure}
\acron implements a secure $\pox$ $=$ $(\Xrequest,$ $\Xexecute,$ $\Xprove,$ $\Xvrfy)$ scheme conforming to Definition~\ref{def:vape}.
The steps in \acron workflow are illustrated in Figure~\ref{fig:seq-events}.
The main idea is to first execute code contained in $ER$.  Then, at some later time, \acron invokes \vrased verified \RA functionality 
to attest the code in $ER$ and include, in the attestation result, additional information that allows \vrf to verify that $ER$ code actually executed.
If $ER$ execution produces an output (e.g., \dev is a sensor running $ER$'s code to obtain some physical/ambient quantity), authenticity and 
integrity of this output can also be verified. That is achieved by including the $EXEC$ flag among inputs to \hmac computed as part of \vrased~\RA.
The value of this flag is controlled by \acron formally verified hardware and \textbf{its memory can not be written by any software running on \dev}.
\acron hardware module runs in parallel with the MCU, monitoring its behavior and deciding the value of $EXEC$ accordingly.

Figure~\ref{fig:vape} depicts \acron's architecture.  In addition to \vrased hardware that provides secure \RA by monitoring a set of 
CPU signals (see Section~\ref{sec:vrased}), \acron monitors values stored in the dedicated physical memory region called $METADATA$.
$METADATA$ contains addresses/pointers to memory boundaries of $ER$ (i.e., $ER_{min}$ and $ER_{max}$) and memory boundaries of 
expected output: $OR_{min}$ and $OR_{max}$. These addresses are sent by \vrf as part of \request, and are configurable at run-time.
The code \xsw to be stored in $ER$ is optionally\footnote{Sending the code to be executed is optional because \xsw might be pre-installed 
on \dev. In that case the proof of execution will allow \vrf to conclude that the pre-installed \xsw was not modified and that it was executed.} 
sent by \vrf.

$METADATA$ includes the $EXEC$ flag, which is initialized to $0$ and only changes from $0$ to $1$ (by \acron's hardware) when $ER$
execution starts, i.e., when the PC points to $ER_{min}$. Afterwards, any violation of \acron's security properties (detailed in 
Section~\ref{sec:properties}) immediately changes $EXEC$ back to $0$. After a violation, the only way to set the flag back to $1$ is to re-start 
execution of $ER$ from the very beginning, i.e., with  PC=$ER_{min}$. In other words, \acron verified hardware makes sure that 
$EXEC$ value covered by the \hmac's result (represented by \token) is $1$, if and only if $ER$ code executed successfully.
As mentioned earlier, we consider an execution to be successful if it runs atomically (i.e., without being interrupted), from its first
$ER_{min}$ to its last instruction $ER_{max}$.

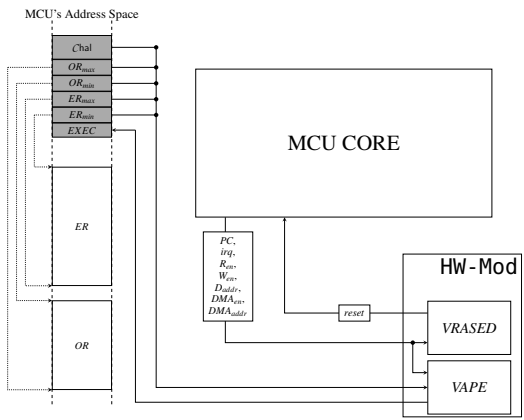
\begin{figure}[!ht]
\centering
\resizebox{.8\columnwidth}{!}{%
	\begin{tikzpicture}[node distance=1.5cm, >=stealth]
\tikzset{
  msp430/.style={
    draw,
    rectangle,
    minimum height=5cm,
    minimum width=10cm,
    align=center
  },
  membb/.style={
    draw,
    rectangle,
    minimum height=9cm,
    minimum width=1.5cm,
    align=center
  },
  mem/.style={
    draw,
    rectangle,
    minimum height=.5cm,
    minimum width=1.5cm,
    align=center
  },
  smallbox/.style={
    draw,
    rectangle,
    minimum height=1.8cm,
    minimum width=2.8cm,
    align=center
  },
  hwmod/.style={
    draw,
    rectangle,
    minimum height=5.5cm,
    minimum width=4cm,
    align=center
  },
  tinybox/.style={
    draw,
    fill={rgb:black,1;white,2},
    rectangle,
    minimum height=.2cm,
    minimum width=2cm,
    align=center
  },
  chalbox/.style={
    draw,
    fill={rgb:black,1;white,2},
    rectangle,
    minimum height=.8cm,
    minimum width=2cm,
    align=center
  },
  erbox/.style={
    draw,
    rectangle,
    minimum height=4cm,
    minimum width=2cm,
    align=center
  },
  orbox/.style={
    draw,
    rectangle,
    minimum height=3cm,
    minimum width=2cm,
    align=center
  }
}
	\node[msp430] (msp) at (0,0) {\huge MCU CORE};
        \node[smallbox, below right = 4cm of msp.south] (vrased) {\Large \vrased};
        \node[smallbox, below = .2cm of vrased.south] (vape) {\Large \acron};
        \node[hwmod] at (4,-6.5) (hwmod) {};
        \node[overlay,anchor=north east] at (hwmod.north east) (test) {\texttt{\huge HW-Mod}};
        
        \path (msp.south) +(-2,0) coordinate (msp_0);
        \path (msp.south) +(-4,0) coordinate (msp_1);
        \path (vrased.west) +(0,.5) coordinate (vrased_0);
        \path (vrased.west) +(0,-.5) coordinate (vrased_1);
        \draw[thick, ->] (msp_1) |- (vrased_1); 
        \node[inner sep=5pt,right, fill=white,draw] at ($(msp_1) + (-.75,-2)$) {\shortstack{$PC$,\\$irq$,\\$R_{en}$,\\$W_{en}$,\\$D_{addr}$,\\$\dmaen$,\\$\dmaaddr$}};
        \draw[thick, <-] (msp_0) |- (vrased_0) ; 
        \node[inner sep=5pt,right, fill=white, draw] at ($(vrased_0) + (-3,0)$) {$reset$};
        
        \node[chalbox, above left = 4cm of msp.west] (chal) {\chal};
        \node[tinybox, below = 0cm of chal.south] (ormax) {$OR_{max}$};
        \node[tinybox, below = 0cm of ormax.south] (ormin) {$OR_{min}$};
        \node[tinybox, below = 0cm of ormin.south] (ermax) {$ER_{max}$};
        \node[tinybox, below = 0cm of ermax.south] (ermin) {$ER_{min}$};
        \node[tinybox, below = 0cm of ermin.south] (exec) {$EXEC$};
        
        \node[erbox, below = 1cm of exec.south] (er) {$ER$};
        \node[orbox, below = .5cm of er.south] (or) {$OR$};
        
        \path (vape.west) +(0,0) coordinate (vape_top);
        \path (vape.west) +(0,-.5) coordinate (vape_bot);
        \path (exec.east) +(.8,0) coordinate (exec_right);
        \draw[thick, -] (vape_bot) -| (exec_right); 
        \draw[thick, ->] (exec_right) -- (exec.east); 
        
        \path (chal.east) +(1.5,0) coordinate (chal_right);
        \path (ormin.east) +(1.5,0) coordinate (ormin_right);
        \path (ormax.east) +(1.5,0) coordinate (ormax_right);
        \path (ermin.east) +(1.5,0) coordinate (ermin_right);
        \path (ermax.east) +(1.5,0) coordinate (ermax_right);
        \draw[thick, <-] (vape_top) -| (chal_right); 
        \draw[thick, -] (chal_right) -- (chal.east); 
        \draw[thick, -] (ormin_right) -- (ormin.east); 
        \draw[thick, -] (ormax_right) -- (ormax.east); 
        \draw[thick, -] (ermin_right) -- (ermin.east); 
        \draw[thick, -] (ermax_right) -- (ermax.east); 
        
		\path (vrased_1) +(-.5, 0) coordinate (vrased_dot);
        \path (vape.west) +(0,0.5) coordinate (vape_connect);
        \draw[thick, ->] (vrased_dot) |- (vape_connect); 
        
        \foreach \n in {chal_right, ormin_right, ormax_right, ermin_right, ermax_right, vrased_dot}
			\node at (\n)[circle,fill,inner sep=1.5pt]{};

		\path (ormax.west) +(-1.5, 0) coordinate (ormax_left);
        \draw[densely dotted, thick, -] (ormax.west) -- (ormax_left); 
        \draw[densely dotted, thick, ->] (ormax_left) |- (or.south west); 

		\path (ormin.west) +(-1.2, 0) coordinate (ormin_left);
        \draw[densely dotted, thick, -] (ormin.west) -- (ormin_left); 
        \draw[densely dotted, thick, ->] (ormin_left) |- (or.north west); 

		\path (ermax.west) +(-.9, 0) coordinate (ermax_left);
        \draw[densely dotted, thick, -] (ermax.west) -- (ermax_left); 
        \draw[densely dotted, thick, ->] (ermax_left) |- (er.south west); 

		\path (ermin.west) +(-.6, 0) coordinate (ermin_left);
        \draw[densely dotted, thick, -] (ermin.west) -- (ermin_left); 
        \draw[densely dotted, thick, ->] (ermin_left) |- (er.north west); 
        
		
        \path (chal.north west) +(0,0.5) coordinate (memleft_start);
        \path (or.south west) +(0,-0.5) coordinate (memleft_end);
        \draw[dashed, thick, -] (memleft_start) -- (memleft_end); 
        \path (chal.north east) +(0,0.5) coordinate (memright_start);
        \path (or.south east) +(0,-0.5) coordinate (memright_end);
        \draw[dashed, thick, -] (memright_start) -- (memright_end); 
        
		\path (chal.north) +(0, 1) coordinate (mem_name);
        \node[overlay,anchor=north] at (mem_name) (test) {\large MCU's Address Space};
        
\end{tikzpicture}
}
\caption{\hw composed of \acron and \vrased hardware modules. Shaded area represents \acron's $METADATA$.}\label{fig:vape}
\end{figure}

In addition to $EXEC$, \hmac covers a set of parameters (in $METADATA$ memory region) that allows \vrf to check whether executed 
software was indeed located in $ER=[ER_{min},ER_{max}]$.
If any output is expected, \vrf specifies a memory range $OR=[OR_{min}, OR_{max}]$ for storing output.
Contents of $OR$ are also covered by the computed \hmac, allowing \vrf to verify authenticity of the output of the execution.

\noindent\textbf{Remark:} Our notion of successful execution requires \xsw to have a single exit point -- $ER_{max}$.
Any self-contained code with multiple legal exits can be trivially instrumented to have a single exit point by 
{\it replacing each exit instruction with a jump to the unified exit point $ER_{max}$.}
This notion also requires \xsw to run atomically. Since this constraint might be undesirable for some real-time systems, 
we discuss how to relax it in Appendix~\ref{apdx:limit}. Finally, \vrf is responsible for defining $OR$ memory 
region according to \xsw behavior. $OR$ should be large enough to fit all output produced by \xsw and $OR$ boundaries 
should correspond to addresses where \xsw writes its output values to be sent to \vrf.

\subsection{\acron's Sub-Properties at a High-Level}\label{sec:properties}
%
We now describe sub-properties enforced by \acron. Section~\ref{sec:verif} formalizes them in LTL 
and provides a single end-to-end definition of \acron correctness. This end-to-end correctness notion is provably implied 
by the composition of all sub-properties. Sub-properties fall into two major groups: \emph{Execution Protection} 
and \emph{Metadata Protection}. A violation of any of these properties implies one or more of:
\begin{compactitem}
\item Code in $ER$ was not executed atomically and in its entirety;
\item Output in $OR$ was not produced by $ER$ execution;
\item Code in $ER$ was not executed in a timely manner, i.e., after receiving the latest \request. 
\end{compactitem}
Whenever \acron detects a violation, $EXEC$ is set to $0$. Since $EXEC$ is included 
among inputs to the computation of \hmac (conveyed in \dev's response), it will be interpreted by \vrf 
as failure to prove execution of code in $ER$.

\noindent\textbf{Remark:} We emphasize that properties discussed below are required 
\textbf{in addition} to \vrased verified properties, i.e., these are entirely different properties used specifically 
to enforce \pox security and should not be viewed as replacements for any of \vrased properties that are used to 
enforce \RA security.

\subsubsection{Execution Protection:}
\textbf{EP1 -- \emph{Ephemeral Immutability:}} Code in $ER$ cannot be modified from the start of its execution until the end 
 	of \sw computation in \prove routine. This property is necessary to ensure that the attestation result reflects 
	the code that executed. Lack of this property would allow \adv to execute some other code $ER_{\adv}$, overwrite 
	it with expected $ER$ and finally call \prove. This would result in a valid proof of execution of $ER$ even though 
	$ER_{\adv}$ was executed instead.
	
\noindent \textbf{EP2 -- \emph{Ephemeral Atomicity:}} $ER$ execution is only considered successful if $ER$ runs starting from $ER_{min}$ 
 	until $ER_{max}$ atomically, i.e., without any interruption. This property conforms with \execute routine in Definition~\ref{def:pox} 
	and with the notion of successful execution in the context of our work. As discussed in Section~\ref{sec:pox_def}, 
	$ER$ must run atomically to prevent malware residing on \dev from interrupting $ER$ execution and resuming it at a different 
	instruction, or modifying intermediate execution results in data memory. Without this property, Return-Oriented Programming (ROP) 
	and similar attacks on $ER$ could change its behavior completely and unpredictably, making any proof of execution
	(and corresponding output) useless.
	
\noindent \textbf{EP3 -- \emph{Output Protection:}} Similar to \textbf{EP1}, \acron must ensure that $OR$ is unmodified from the time after 
        $ER$ code execution is finished until completion of \hmac computation in \prove. Lack of this property would allow \adv 
        to overwrite $OR$ and successfully spoof $OR$ produced by $ER$, thus convincing \vrf that it produced output $OR^{\adv}$.

\subsubsection{Metadata Protection:}
%
\textbf{MP1 - \emph{Executable/Output ($ER$/$OR$) Boundaries:}} \acron hardware ensures properties \textbf{EP1}, \textbf{EP2}, and \textbf{EP3} 
 	according to values: $ER_{min}$, $ER_{max}$, $OR_{min}$, $OR_{max}$. These values are configurable and can be decided by \vrf 
	based on application needs. They are written into metadata-dedicated physical addresses in \dev memory before $ER$ execution.
 	Therefore, once $ER$ execution starts, \acron hardware must ensure that such values remain unchanged until \prove completes. 
	Otherwise, \adv could generate valid attestation results, by attesting [$ER_{min}$, $ER_{max}$], while, in fact, having executed code in 
	a different region: [$ER_{min}^{\adv}$, $ER_{max}^{\adv}$].
	
\noindent \textbf{MP2 - \emph{Response Protection:}} The appropriate response to \vrf's challenge must be unforgeable and non-invertible. 
        Therefore, in the \prove routine, \attkey used to compute \hmac must never be leaked (with non-negligible probability) and 
        \hmac implementation must be functionally correct, i.e., adhere to its cryptographic specification. Moreover, contents of memory 
        being attested must not change during \hmac computation. We rely on \vrased to ensure these properties. Also, to ensure 
        trustworthiness of the response, \acron guarantees that no software in \dev can ever modify $EXEC$ flag and that, once $EXEC = 0$, 
        it can only become $1$ if $ER$'s execution re-starts afresh.
	
\noindent \textbf{MP3 - \emph{Challenge Temporal Consistency:}} \acron must ensure that $\chal$ cannot be modified between $ER$'s 
	execution and \hmac computation in \prove. Without this property, the following attack is possible: (1) \dev-resident malware 
	executes $ER$ properly (i.e., by not violating \textbf{EP1-EP3} and \textbf{MP1-MP2}), resulting in $EXEC=1$ after execution 
	stops, and (2) at a later time, malware receives \chal from \vrf and simply calls \prove on this \chal without executing $ER$. 
	As a result, malware would acquire a valid proof of execution (since $EXEC$ remains $1$ when the proof is generated) 
	even though no $ER$ execution occurred before \chal was received.
 	Such attacks are prevented by setting $EXEC=0$ whenever the memory region storing $\chal$ is modified.

\section{Formal Specification \& Verified Implementation}\label{sec:verif}
Our formal verification approach starts by formalizing \acron sub-properties Linear 
Temporal Logic (LTL) to define invariants that must hold throughout the MCU operation. We then use a 
theorem prover~\cite{spot} to write a computer-aided proof that the conjunction of the LTL sub-properties imply an 
end-to-end formal definition for the guarantee expected from \acron hardware. \acron correctness, when properly 
composed with \vrased guarantees, yields a \pox scheme secure according to Definition~\ref{def:pox_sec_def}.
This is proved by showing that, if the composition between the two is implemented as described in Definition~\ref{def:vape}, 
\vrased security can be reduced to \acron security.

\acron hardware module is composed of several sub-modules written in Verilog Hardware Description Language (HDL).
Each sub-module is responsible for enforcing a set of LTL sub-properties and is described as an FSM in Verilog at Register Transfer Level (RTL).
Individual sub-modules are combined into a single Verilog design. The resulting composition is converted to the SMV model checking 
language using the automatic translation tool Verilog2SMV~\cite{irfan2016verilog2smv}. The resulting SMV is simultaneously verified against 
all LTL specifications, using the model checker NuSMV\cite{nusmv}, to prove that the final Verilog of \acron complies with all necessary properties.

\subsection{Machine Model}\label{sec:model}
Definition~\ref{def:exec_model} models, in LTL, the behavior of low-end MCUs considered in this work.
It consists of a subset of the machine model introduced by \vrased.
Nonetheless, this subset models all MCU behavior relevant for stating and verifying correctness of \acron's implementation.

\begin{figure}[!ht]
\begin{mdframed}
\begin{definition}{Machine Model (subset)} \label{def:exec_model}
\begin{compactenum}~
\footnotesize
 \item Modify\_Mem($i$) $\rightarrow (W_{en} \land D_{addr} = i) \lor (DMA_{en} \land DMA_{addr} = i)$
 \item Interrupt $\rightarrow$ $irq$
 \item $MR$, $CR$, $AR$, $KR$, $XS$, and $METADATA$ are non-overlapping memory regions
\end{compactenum}
\end{definition}
\end{mdframed}
\end{figure}

\texttt{Modify\_Mem} models that a given memory address can be modified by a CPU 
instruction or by a DMA access. In the former, $W_{en}$ signal must be set and $D_{addr}$ must contain the 
target memory address. In the latter, $DMA_{en}$ signal must be set and $DMA_{addr}$ 
must contain the target DMA address. The requirements for {\em reading from} a memory address are 
similar, except that instead of $W_{en}$, $R_{en}$ must be on. We do not explicitly state this behavior since 
it is not used in \acron proofs.  For the same reason, modeling the effects of instructions that only modify 
register values (e.g., ALU operations, such as $add$ and $mul$) is also not necessary.  The machine model 
also captures the fact that, when an interrupt happens during execution, the $irq$ signal in MCU hardware is set to $1$.

With respect to memory layout, the model states that $MR$, $CR$, $AR$, $KR$, $XS$, and $METADATA$ 
are disjoint memory regions. The first five memory regions are defined in \vrased. 
As shown in Figure~\ref{fig:vape}, $METADATA$ is a fixed memory region used by \acron to store information about 
software execution status.

\subsection{Security \& Implementation Correctness}
We use a two-part strategy to prove that \acron is a secure \pox architecture, according to Definition~\ref{def:pox_sec_def}:
\begin{compactenum}
 \item[{\bf [A]:}] We show that properties \textbf{EP1-EP3} and \textbf{MP1-MP3}, discussed in Section~\ref{sec:properties} and 
 	formally specified next in Section~\ref{sec:sub_prop_ltl}, are sufficient to guarantee that $EXEC$ flag is $1$ iff \xsw indeed 
	executed on \dev. To show this, we compose a computer proof using SPOT LTL proof assistant~\cite{spot}.
\item[{\bf [B]:}] We use cryptographic reduction proofs to show that, as long as part {\bf A} holds, \vrased security 
	can be reduced to \acron's \pox security from Definition~\ref{def:pox_sec_def}.
 	In turn, \hmac's existential unforgeability can be reduced to \vrased's security~\cite{vrasedp}. Therefore, both \acron and 
	\vrased rely on the assumption that \hmac is a secure \texttt{MAC}.
\end{compactenum}

\begin{figure*}[!ht]
\begin{mdframed}
\begin{definition}\label{def:vape_fc} Formal specification of \acron's correctness.
\footnotesize
 \begin{align*}
 \begin{split}
 & \{ \\
 & \quad PC = ER_{min} ~ \land ~ [(PC \in ER \land \neg Interrupt \land \neg reset \land \neg DMA_{en}) \quad \textbf{U} \quad PC = ER_{max}] \quad \land \\
 & \quad [(\neg~\text{Modify\_Mem}(ER) \land \neg~\text{Modify\_Mem}(METADATA) \land (PC \in ER \lor \neg~\text{Modify\_Mem}(OR))) \quad \textbf{U} \quad PC = CR_{min}] \quad \\
 & \} \quad \textbf{B} \quad \{EXEC \land PC \in CR\}
 \end{split}
 \end{align*}
\end{definition}
\end{mdframed}
\vspace{-0.5cm}
\begin{mdframed}
\begin{definition}{Necessary Sub-Properties for Secure Proofs of Execution in LTL.}\label{def:LTL_props}~\\
\footnotesize
Ephemeral Immutability:
\begin{equation}\label{eq:ephe_immut}
\begin{split}
\text{\bf G}: \ \{
[W_{en} \land (D_{addr} \in ER)] \lor [DMA_{en} \land (DMA_{addr} \in ER)] \rightarrow \neg EXEC \}
\end{split}
\end{equation}

Ephemeral Atomicity:
\begin{equation}\label{eq:ephe_atom1}
\begin{split}
\text{\bf G}: \ \{(PC \in ER) \land \neg (\text{\bf X}(PC) \in ER) \rightarrow PC = ER_{max} \lor \neg \text{\bf X}(EXEC) \ \}
\end{split}
\end{equation}

\begin{equation}\label{eq:ephe_atom2}
\begin{split}
 & \text{\bf G}: \ \{\neg (PC \in ER) \land (\text{\bf X}(PC) \in ER) \rightarrow \text{\bf X}(PC) = ER_{min} \lor \neg \text{\bf X}(EXEC) \}
\end{split}
\end{equation}

\begin{equation}\label{eq:ephe_atom3}
\begin{split}
 & \text{\bf G}: \ \{(PC \in ER) \land irq \rightarrow \neg EXEC \}
\end{split}
\end{equation}

Output Protection:
\begin{equation}\label{eq:output_prot}
\begin{split}
\text{\bf G}: \ \{
[\neg(PC \in ER) \land (W_{en} \land D_{addr} \in OR)] \lor (DMA_{en} \land DMA_{addr} \in OR) \lor (PC \in ER \land DMA_{en}) \rightarrow \neg EXEC \}
\end{split}
\end{equation}

Executable/Output (ER/OR) Boundaries \& Challenge Temporal Consistency:
\begin{equation}\label{eq:boundaries}
\begin{split}
 & \text{\bf G}: \ \{ER_{min} > ER_{max} \lor OR_{min} > OR_{max}\rightarrow \neg EXEC \}
\end{split}
\end{equation}

\begin{equation}\label{eq:no_overlap}
\begin{split}
 & \text{\bf G}: \ \{ER_{min} \leq CR_{max} \lor ER_{max} > CR_{max} \rightarrow \neg EXEC \}
\end{split}
\end{equation}

\begin{equation}\label{eq:metadata_prot}
\begin{split}
\text{\bf G}: \ \{
[W_{en} \land (D_{addr} \in METADATA)] \lor [DMA_{en} \land (DMA_{addr} \in METADATA)] \rightarrow \neg EXEC \}
\end{split}
\end{equation}
\begin{center}
\textbf{Remark:} Note that $\chal_{mem} \in METADATA$. 
\end{center}

Response Protection:
\begin{equation}\label{eq:EXEC_ZERO}
\begin{split}
\text{\bf G}: \ \{
\neg EXEC \land \text{\bf X}(EXEC) \rightarrow \text{\bf X}(PC = ER_{min})\}
\end{split}
\end{equation}

\begin{equation}\label{eq:reset_prot}
\begin{split}
\text{\bf G}: \ \{
 reset \rightarrow \neg EXEC \}
\end{split}
\end{equation}

\end{definition}
\end{mdframed}
\end{figure*}

\begin{figure}[!ht]
 \hspace*{-1.5cm}
\centering
\resizebox{.85\columnwidth}{!}{\begin{tikzpicture}[node distance=1.5cm, >=stealth]
\tikzset{
  shbox/.style={
    draw,
    rectangle,
    fill={rgb:black,1;white,2}, 
    pattern=north west lines,
    minimum height=.5cm,
    minimum width=.5cm,
    align=center
  },
  obox/.style={
    draw,
    rectangle,
    fill={rgb:black,1;white,2}, 
    minimum height=.5cm,
    minimum width=.5cm,
    align=center
  }
}

	\coordinate (origin) at (0,0);
	\coordinate (xmax) at (8.5,0);
	\coordinate (ymax) at (0,4.7);
	\draw[thick, ->] (origin) -- (xmax) ;  
	
	\foreach \x in {0.5,1.5,3.5,5,6.5,7.5}
    	\draw (\x cm,3pt) -- (\x cm,-3pt);
      
    \draw (0.5,0) node[below=3pt] {\scriptsize $t_{req}$};
    \draw (1.5,0) node[below=3pt] {\scriptsize $t(ER_{min})$};
    \draw (3.5,0) node[below=3pt] {\scriptsize $t(ER_{max})$};
    \draw (5,0) node[below=3pt] {\scriptsize $t(CR_{min})$};
    \draw (6.5,0) node[below=3pt] {\scriptsize $t(CR_{max})$};
    \draw (7.5,0) node[below=3pt] {\scriptsize $t_{verif}$};
    \draw (8.5,0) node[below=3pt] {Time};

	\draw[thick, ->] (origin) -- (ymax);
	
	\foreach \y in {1,2,3}
    	\draw (3pt, \y cm) -- (-3pt, \y cm);
    \draw (0,1) node[left=3pt] {\small $OR$};
    \draw (0,2) node[left=3pt] {\small $ER$};
    \draw (0,3) node[left=3pt] {\small \shortstack{$META$\\$DATA$}};
    \draw (0,4.5) node[left=3pt] {Region};
    
    \draw [fill={rgb:black,1;white,2}, pattern=north west lines] (1.5,2.7) rectangle (5,3.3);
    \draw [fill={rgb:black,1;white,2}] (5,2.7) rectangle (6.5,3.3);
    
    \draw [fill={rgb:black,1;white,2}, pattern=north west lines] (1.5,1.7) rectangle (5,2.3);
    \draw [fill={rgb:black,1;white,2}] (5,1.7) rectangle (6.5,2.3);
    
    \draw [fill={rgb:black,1;white,2}, pattern=north west lines] (3.5,0.7) rectangle (5,1.3);
    \draw [fill={rgb:black,1;white,2}] (5,0.7) rectangle (6.5,1.3);
    
    \draw [dashed] (1.5,0) -- (1.5,4.5);
    \draw (1.5,4.5) node[above=3pt] {State $S_0$};
    \draw [dashed] (3.5,0) -- (3.5,4.5);
    \draw (3.5,4.5) node[above=3pt] {State $S_1$};
    \draw [dashed] (5,0) -- (5,4.5);
    \draw (5,4.5) node[above=3pt] {State $S_2$};
    \draw [dashed] (6.5,0) -- (6.5,4.5);
    \draw (6.5,4.5) node[above=3pt] {{\ensuremath{\sf{\mathcal H}}}\xspace ready};
    
    \draw [decorate,decoration={brace,amplitude=10pt,raise=0pt},yshift=0pt] (1.5,3.7) -- (3.5,3.7) node [black,midway,yshift=.5cm] {$ER$ execution};
    \draw [decorate,decoration={brace,amplitude=10pt,raise=0pt},yshift=0pt] (5,3.7) -- (6.5,3.7) node [black,midway,yshift=.5cm] {Attestation};
    
    \node [shbox] (legend1) at (7.5,3) {};
        \node[overlay,anchor=west] at (legend1.east) (test) {\small\shortstack{Unchanged memory\\required by \acron}};
    \node [obox] (legend2) at (7.5, 2) {};
        \node[overlay,anchor=west] at (legend2.east) (test) {\small\shortstack{Unchanged memory\\enforced by \vrased}};

\end{tikzpicture}}
\caption{Illustration of time intervals that each memory region must remain unchanged in order to produce a valid \token ($EXEC = 1$).
$t(X)$ denotes the time when $PC=X$.}\label{fig:memconsist}
\end{figure}
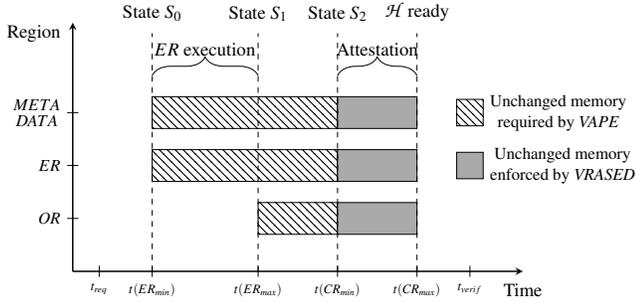

In the rest of this section, we convey the intuition behind both of these steps. Proof details are in Appendix~\ref{apdx:prove}.

The goal of part {\bf  A} is to show that \acron's sub-properties imply Definition~\ref{def:vape_fc}.
LTL specification in Definition~\ref{def:vape_fc} captures the conditions that must hold in order for  
$EXEC$ to be set to 1 during execution of \prove, enabling generation of a valid proof of execution.
This specification ensures that, in order to have $EXEC=1$ during execution of \prove (i.e, for 
$[EXEC \land PC \in CR]$ to hold), at least once \textbf{before such time} the following must have happened:
\begin{compactenum}
 \item The system reached state $S_0$ where software stored in $ER$ started executing from its first 
 	instruction ($PC = ER_{min}$).
 \item The system eventually reached a state $S_1$ when ER finished executing ($PC = ER_{max}$). 
 	In the interval between $S_0$ and $S_1$ $PC$ kept executing instructions within $ER$, there 
	were no interrupts, no resets, and DMA remained inactive.
 \item The system eventually reached a state $S_2$ when \prove started executing ($PC = CR_{min}$). 
 	In the interval between $S_0$ and $S_2$,  $METADATA$ and $ER$ regions were not modified.
 \item In the interval between $S_0$ and $S_2$,  $OR$ region was only modified by $ER$'s execution,
 	i.e., $PC \in ER \lor \neg~\text{Modify\_Mem}(OR)$.
\end{compactenum}
Figure~\ref{fig:memconsist} shows the time windows wherein each memory region must not change 
during \acron's \pox as implied by \acron's correctness (Definition~\ref{def:vape_fc}).
Violating any of these conditions will cause $EXEC$ have value $0$ during \prove's computation.
Consequently, any violation will result in \vrf rejecting the proof of execution since it will not conform to 
the expected value of \token, per Equation~\ref{eq:expected_token} in Definition~\ref{def:vape}.

The intuition behind the cryptographic reduction (part {\bf B}) is that computing \token involves simply 
invoking \vrased \sw with $MR=\chal$, $ER \in AR$, $OR \in AR$, and $METADATA \in AR$.
Therefore, a successful forgery of \acron's \token implies breaking \vrased security.  Since \token always 
includes the value of $EXEC$, this implies that \acron is \pox-secure (Definition~\ref{def:pox_sec_def}).
The complete reduction is presented in Appendix~\ref{apdx:prove}.

\subsection{\acron's Sub-Properties in LTL}\label{sec:sub_prop_ltl}
We formalize the necessary sub-properties enforced by \acron as LTL 
specifications~\ref{eq:ephe_immut}--\ref{eq:reset_prot} in Definition~\ref{def:LTL_props}. We describe 
how they map to high-level notions \textbf{EP1-EP3} and \textbf{MP1-MP3} discussed in Section~\ref{sec:properties}.
Appendix~\ref{apdx:prove} discusses a computer proof that the conjunction of this set of properties is sufficient to satisfy a 
formal definition of \acron correctness from Definition~\ref{def:vape_fc}. 
  
LTL~\ref{eq:ephe_immut} enforces \textbf{EP1 -- Ephemeral immutability} by making sure that whenever $ER$ 
memory region is written by either CPU or DMA, $EXEC$ is immediately set to logical 0 (false).

\textbf{EP2 -- Ephemeral Atomicity} is enforced by a set of three LTL specifications. LTL~\ref{eq:ephe_atom1}
enforces that the only way for $ER$'s execution to terminate, without setting $EXEC$ to logical $0$, is through its last 
instruction: $PC = ER_{max}$. This is specified by checking the relation between current and next $PC$ values using 
LTL ne\textbf{X}t operator. In particular, if current $PC$ value is within $ER$, and next $PC$ value is outside \sw region, 
then either current $PC$ value is the address of $ER_{max}$, or $EXEC$ is set to $0$ in the next cycle.
Also, LTL~\ref{eq:ephe_atom2} enforces that the only way for $PC$ to enter $ER$ is through the very first 
instruction: $ER_{min}$. This prevents $ER$ execution from starting at  some point in the middle of $ER$, thus making 
sure that $ER$ always executes in its entirety.  Finally, LTL~\ref{eq:ephe_atom3} enforces that $EXEC$ is 
set to zero if an interrupt happens in the middle of $ER$ execution. Even though LTLs~\ref{eq:ephe_atom1} and~\ref{eq:ephe_atom2} 
already enforce that PC cannot change to anywhere outside $ER$, interrupts could be programmed to return to an arbitrary 
instruction within $ER$. Although this would not violate LTLs~\ref{eq:ephe_atom1} and~\ref{eq:ephe_atom2},  it would still modify 
$ER$'s behavior. Therefore, LTL~\ref{eq:ephe_atom3} is needed to prevent that.

\textbf{EP3 -- Output Protection} is enforced by LTL~\ref{eq:output_prot} by making sure that: (1) DMA controller does not write 
into  $OR$; (2) CPU can only modify $OR$ when executing instructions within $ER$;  and 3) DMA cannot be active during 
$ER$ execution; otherwise, a compromised DMA could change intermediate results of $ER$ computation in data memory, 
potentially modifying $ER$ behavior.

Similar to \textbf{EP3}, \textbf{MP1 -- Executable/Output Boundaries} and \textbf{MP3 -- Challenge Temporal Consistency} are enforced 
by LTL~\ref{eq:metadata_prot}.  Since \chal as well as  $ER_{min}$, $ER_{max}$, $OR_{min}$, and $OR_{max}$ are all stored in 
$METADATA$ reserved memory region, it suffices to ensure that $EXEC$ is set to logical $0$ whenever this region is modified.
Also, LTL~\ref{eq:boundaries} enforces that $EXEC$ is only set to one if $ER$ and $OR$ are configured (by $METADATA$ 
values $ER_{min}$, $ER_{max}$, $OR_{min}$, $OR_{max}$) as valid memory regions.

Finally, LTLs~\ref{eq:EXEC_ZERO}, and~\ref{eq:reset_prot} (in addition to \vrased verified \RA architecture) are responsible for 
ensuring \textbf{MP2- Response Protection} by making sure that $EXEC$ always reflects what is intended by \acron hardware.
LTL~\ref{eq:output_prot} specifies that the only way to change $EXEC$ from $0$ to $1$ is by starting $ER$'s execution over.
Finally, LTL~\ref{eq:reset_prot} states that, whenever a reset happens (this also includes the system initial booting state) and 
execution is initialized, the initial value of $EXEC$ is $0$.

To conclude, recall that $EXEC$ is read-only to all software running on \dev. Therefore, malware can not change it directly.

\acron is designed as a set of seven hardware sub-modules, each verified to enforce a subset of properties discussed in 
this section. Due to space constraints, examples of implementation of verified sub-modules as FSMs are discussed 
in Appendix~\ref{apdx:sub-modules}.

\section{Implementation \& Evaluation}\label{sec:eval}
\ignore{
In summary, \acron incurs modest hardware overhead, compared to the \vrased baseline:  $\approx 2\%$ for registers and $12\%$ for LUTs.
The runtime to produce a proof of \xsw execution depends on the size of \xsw, which determines \vrased's attestation runtime 
used to produce \token. In the most expensive or extreme case, when the entire program memory ($8$ $kB$) is occupied by $ER+OR$, 
this computation takes around $900$ms on the 8MHz MSP430.
Through the rest of this section we discuss \acron implementation and evaluation in more details.
Finally, as a proof of concept, Section~\ref{sec:auth_sensing} discusses how the \pox, functionality realized in a provably secure manner by \acron, can be used to implement real low-cost sensors and actuators that ``cannot lie'' about their produced outputs, even under the assumption of full compromise of their software state.
}
\acron implementation uses OpenMSP430~\cite{openmsp430} as its open core implementation.
We implement the hardware architecture shown in Figure~\ref{fig:vape}. In addition to \acron and \vrased modules in 
\hw, we implement a peripheral module responsible for storing and maintaining \acron $METADATA$.
As a peripheral, contents of $METADATA$ can be accessed in a pre-defined memory address via 
standard peripheral memory access. We also ensure that $EXEC$ (located inside $METADATA$) is 
unmodifiable in software by removing software-write wires in hardware. Finally, we use Xilinx Vivado to synthesize
an RTL description of the modified \hw and deploy it on the Artix-7 FPGA class.

\subsection{Evaluation Results}
\begin{table*}[!h]
\footnotesize
\centering
\begin{tabular}{l|cc|c|cccc}
\hline
                                             & \multicolumn{2}{c|}{Hardware} & Reserved & \multicolumn{4}{c}{Verification} \\
                                             & Reg & LUT & \ram (byte) & \# LTL Invariants & Verified Verilog LoC & Time (s) & Mem (MB) \\ \hline\hline
\multicolumn{1}{l|}{OpenMSP430~\cite{openmsp430}}             &  691   & 1904 & 0 & - & - & - & -   \\
\multicolumn{1}{l|}{\vrased~\cite{vrasedp}}  &  721   &  1964 & 2332 & 10 & 481 & 0.4 & 13.6  \\
\multicolumn{1}{l|}{\acron+\vrased}   & 735   &  2206 & 2341 & 20 & 1385 & 183.6 & 280.3  \\ \hline
\end{tabular}%
\caption{Evaluation results.}
\label{tab:hw}
\end{table*}

\noindent\textbf{Hardware \& Memory Overhead.} 
Table~\ref{tab:hw} reports \acron hardware overhead as compared to unmodified OpenMSP430~\cite{openmsp430} and 
\vrased~\cite{vrasedp}. \acron hardware overhead is small compared to the baseline \vrased; it requires 2\% and 12\% 
additional registers and LUTs, respectively. In absolute numbers, it adds 44 registers and 302 Look-Up Tables (LUTs) 
to the underlying MCU. In terms of memory, \acron needs 9 extra bytes of \ram for storing $METADATA$.
This overhead corresponds to 0.01\% of MSP430 16-bit address space.

\noindent\textbf{Run-time.}
We do not observe any overhead for software's execution time on the \acron-enabled \dev since \acron does not introduce
new instructions or modifications to the MSP430 ISA. \acron hardware runs in parallel with the original MSP430 CPU.
Run-time to produce a proof of \xsw execution includes: (1) time to execute \xsw (\execute), and (2) time to compute an 
attestation token (\prove). The former only depends on \xsw behavior itself (e.g., \sw can be a small sequence of 
instructions or have long loops). As mentioned earlier, \acron does not affect \xsw runtime.
\prove's run-time is linear in the size of $ER+OR$. In the worst-case scenario where these regions occupy the entire program 
$8$kB memory, \prove takes around $900$ms on an 8MHz device.

\noindent\textbf{Verification Efforts.} 
We verify \acron on an Ubuntu 16.04 machine running at 3.40GHz. Results are shown in Table~\ref{tab:hw}.
\acron verification requires checking 10 extra invariants (shown in Definition~\ref{def:LTL_props}) in addition to 
existing \vrased invariants. It also consumes significantly higher run-time and memory usage than \vrased verification.
This is because additional invariants introduce five additional variables ($ER_{min}$, $ER_{max}$, $OR_{min}$, $OR_{max}$ and $EXEC$),
potentially resulting in an exponential increase in complexity of the model checking process. Nonetheless, the overall verification process is 
still reasonable for a commodity desktop -- it takes around 3 minutes and consumes 280MB of memory.

\subsection{Comparison with CFA}\label{sec:experiments}
To the best of our knowledge, \acron is the first of its kind and thus there are no other directly comparable \pox architectures.
However, to provide a (performance and overhead) point of reference and a comparison, we contrast \acron overhead with 
that state-of-the-art CFA architectures. As discussed in Section~\ref{sec:RW}, even though CFA is not 
directly applicable for producing proofs of execution with authenticated outputs, we consider it to be the closest-related service, 
since it reports on the exact execution path of a program.

We consider three recent CFA architectures: Atrium~\cite{zeitouni2017atrium}, LiteHAX~\cite{dessouky2018litehax}, and 
LO-FAT~\cite{dessouky2017fat}. Figure~\ref{fig:comparison}.a compares \acron to these architectures in terms of number 
of additional LUTs. In this figure, the black dashed line represents the total cost of the MSP430 MCU: 1904 LUTs.
Figure~\ref{fig:comparison}.b presents a similar comparison for the amount of additional registers required by these architectures.
In this case, the total cost of the MSP430 MCU itself is of 691 registers.
Finally, Figure~\ref{fig:comparison}.c presents the amount of dedicated RAM required by these architectures
(\acron's dedicated RAM corresponds to the exclusive access stack implemented by \vrased).

As expected, \acron incurs much lower overhead. According to our results, the cheapest CFA architecture, LiteHAX, would entail
an overhead of nearly 100\% LUTs and 300\% registers, on MSP430. In addition, LiteHAX would require $150$ $kB$ of dedicated RAM. 
This amount far exceeds entire addressable memory ($64$ $kB$) of 16-bit processors, such as MSP430.
Results support our claim that CFA is not applicable to this class of low-end devices.
Meanwhile, \acron needs a total of 12\% additional LUTs and 2\% additional registers.
\vrased requires about $2$ $kB$ of reserved RAM, which is not increased by \acron \pox support.

\begin{figure}[t]
	\centering
	\subfigure[Additional HW overhead (\%) in Number of Look-Up Tables]
	{\includegraphics[width=0.45\columnwidth]{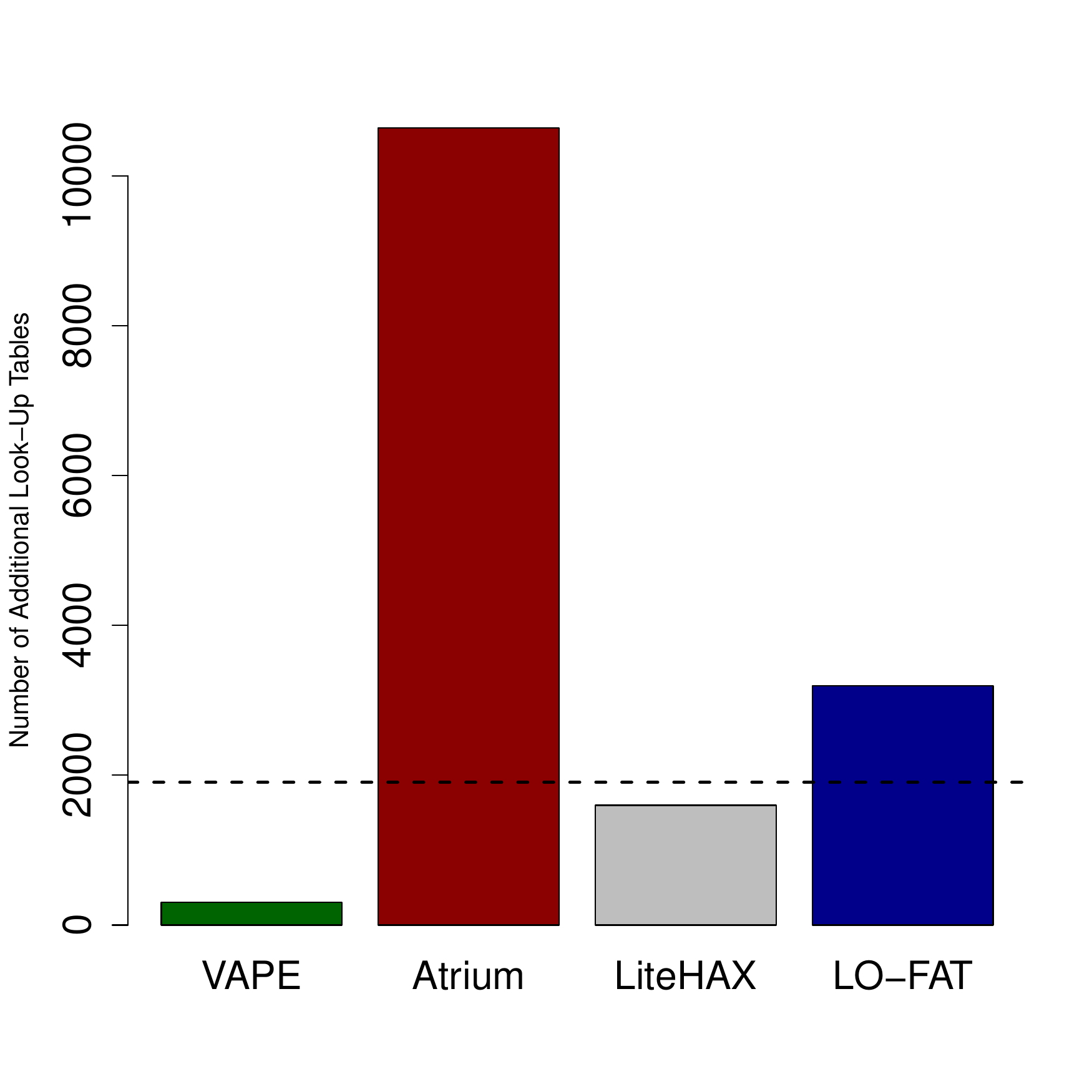}}
	\subfigure[Additional HW overhead (\%) in Number of Registers]
	{\includegraphics[width=0.45\columnwidth]{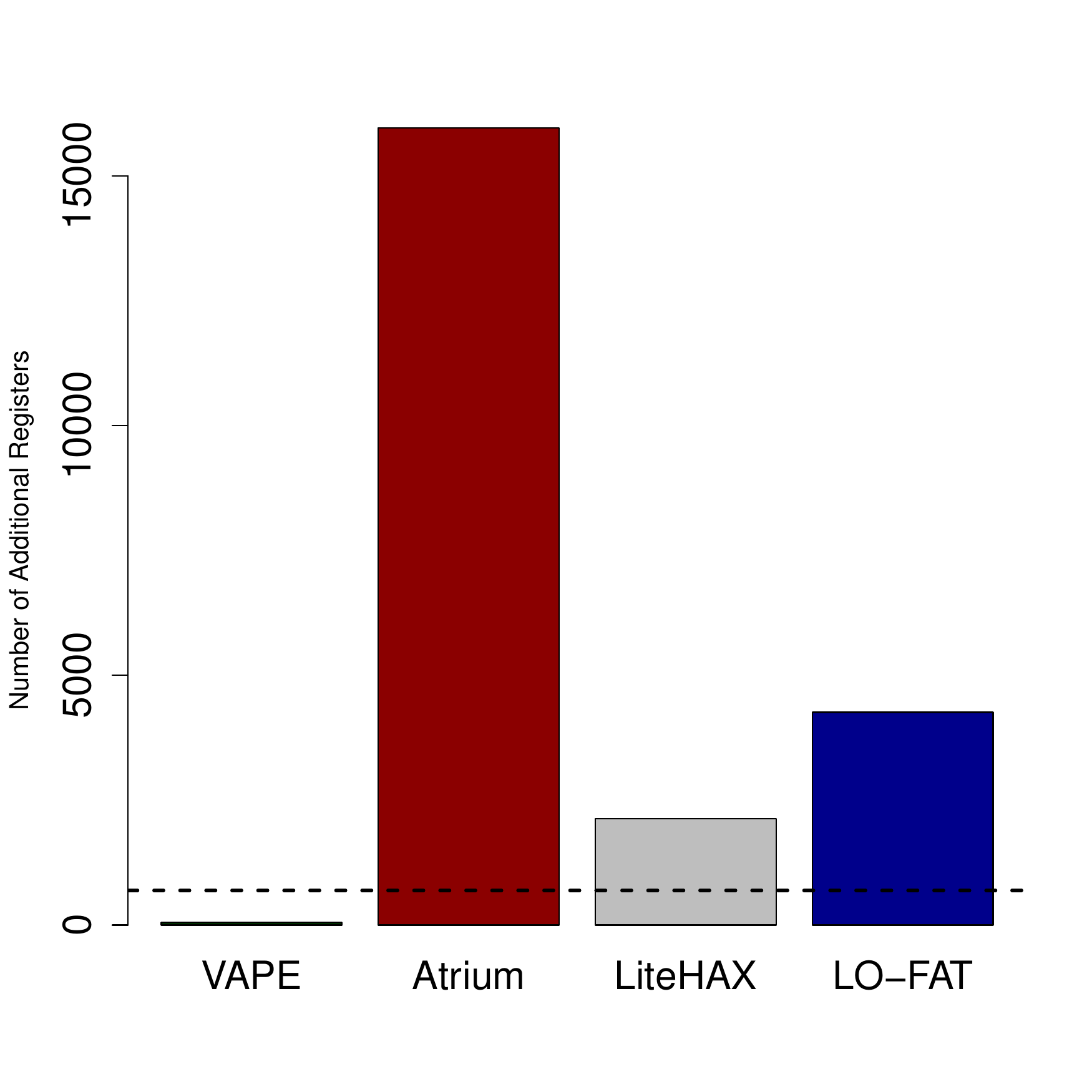}}
	\subfigure[Dedicated RAM]
	{\includegraphics[width=0.45\columnwidth]{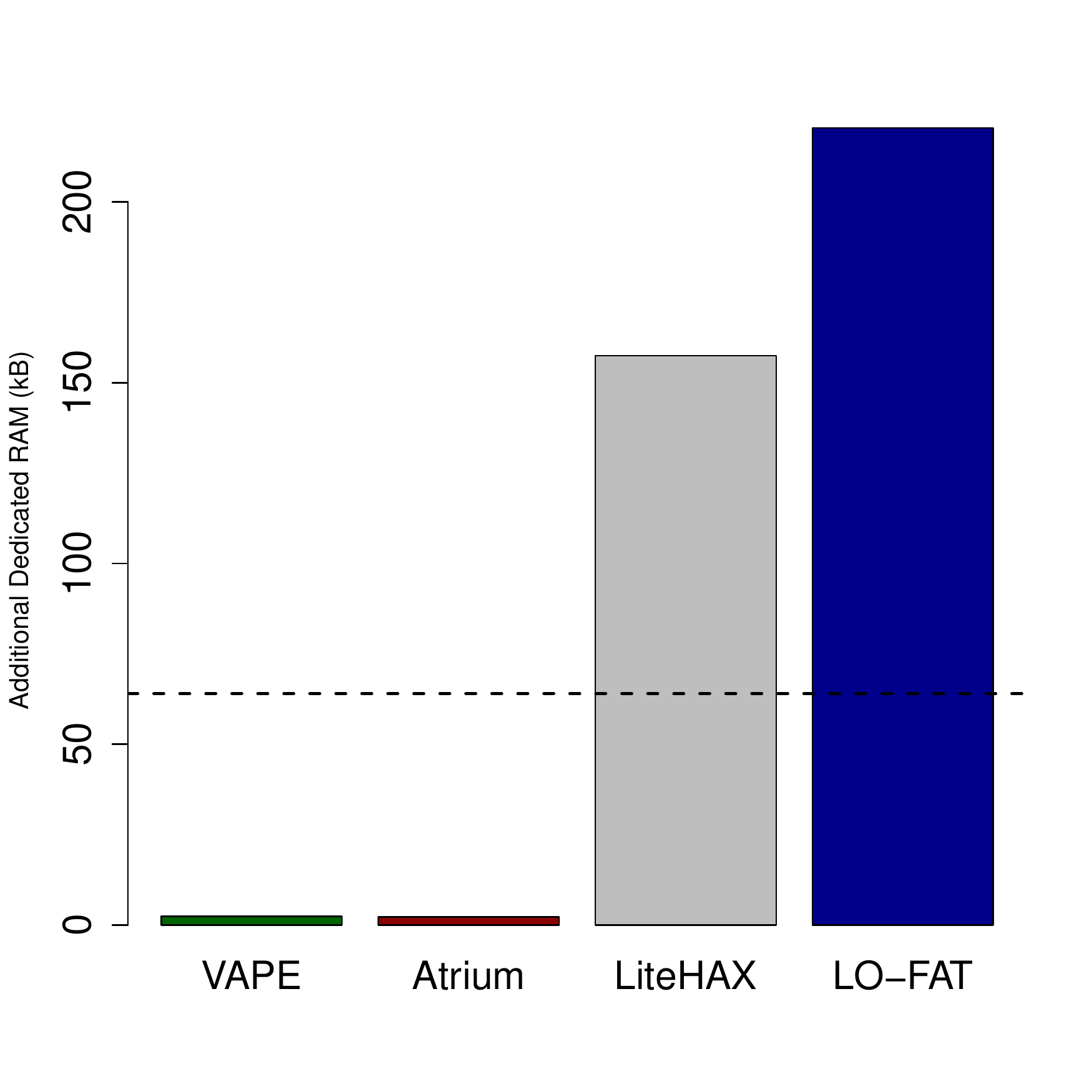}}
	\caption{Overhead comparison between \acron and CFA architectures. Dashed lines in (a) and (b)represent the total 
	hardware cost of MSP430. Dashed line in (c) represents total addressable memory ($64$ $kB$) on MSP430.}\label{fig:comparison}
	\vspace{-2em}
\end{figure}

\subsection{Proof of Concept: Authenticated Sensing and Actuation}\label{sec:auth_sensing}
As discussed in Section~\ref{intro} an important functionality attainable with \pox is authenticated sensing/actuation. 
In this section, we demonstrate how to use \acron to build sensors and actuators that ``cannot lie''.

As a running example we use a fire sensor: a safety-critical low-end embedded device commonly 
present in households and workplaces. It consists of an MCU equipped with analog 
hardware for measuring physical/chemical quantities, e.g., temperature, humidity, and $CO_2$ level.
It is also usually equipped with actuation-capable analog hardware, such as a buzzer.
Analog hardware components are directly connected to MCU General Purpose Input/Output (GPIO) ports.
GPIO ports are physical wires directly mapped to fixed memory locations in MCU memory.
Therefore, software running on the MCU can read physical quantities directly from GPIO memory.

In this example, we consider that MCU software periodically reads these values and transmits them to 
a remote safety authority, e.g., a fire department, which then decides whether to take action.
The MCU also triggers the buzzer actuator whenever sensed values indicate a fire.
Given the safety-critical nature of this application, the safety authority must be assured
that reported values are authentic and were produced by execution of expected software.
Otherwise, malware could spoof such values (e.g., by not reading them from the proper GPIO).
\pox guarantees that reported values were read from the correct GPIO port 
(since the memory address is specified by instructions in the ER executable), and
produced output (stored in OR) was indeed generated by execution of ER and was unmodified thereafter.
Thus, upon receiving sensed values accompanied by a \pox, the safety authority is assured 
that the reported sensed value can be trusted.

As a proof of concept, we use \acron to implement a simple fire sensor that operates with temperature and humidity quantities. 
It communicates with a remote \vrf (e.g., the fire department) using a low-power ZigBee radio\footnote{https://www.zigbee.org/} 
typically used by low-end CPS/IoT devices. Temperature and humidity analog devices are connected to a 
\acron-enabled MSP430 MCU running at 8MHz and synthesized using a Basys3 Artix-7 FPGA board.
As shown in Figure~\ref{fig:fire}, MCU GPIO ports connected to the temperature/humidity sensor and to the buzzer.
\begin{figure}[!h]
	\centering
	\includegraphics[width=.8\linewidth]{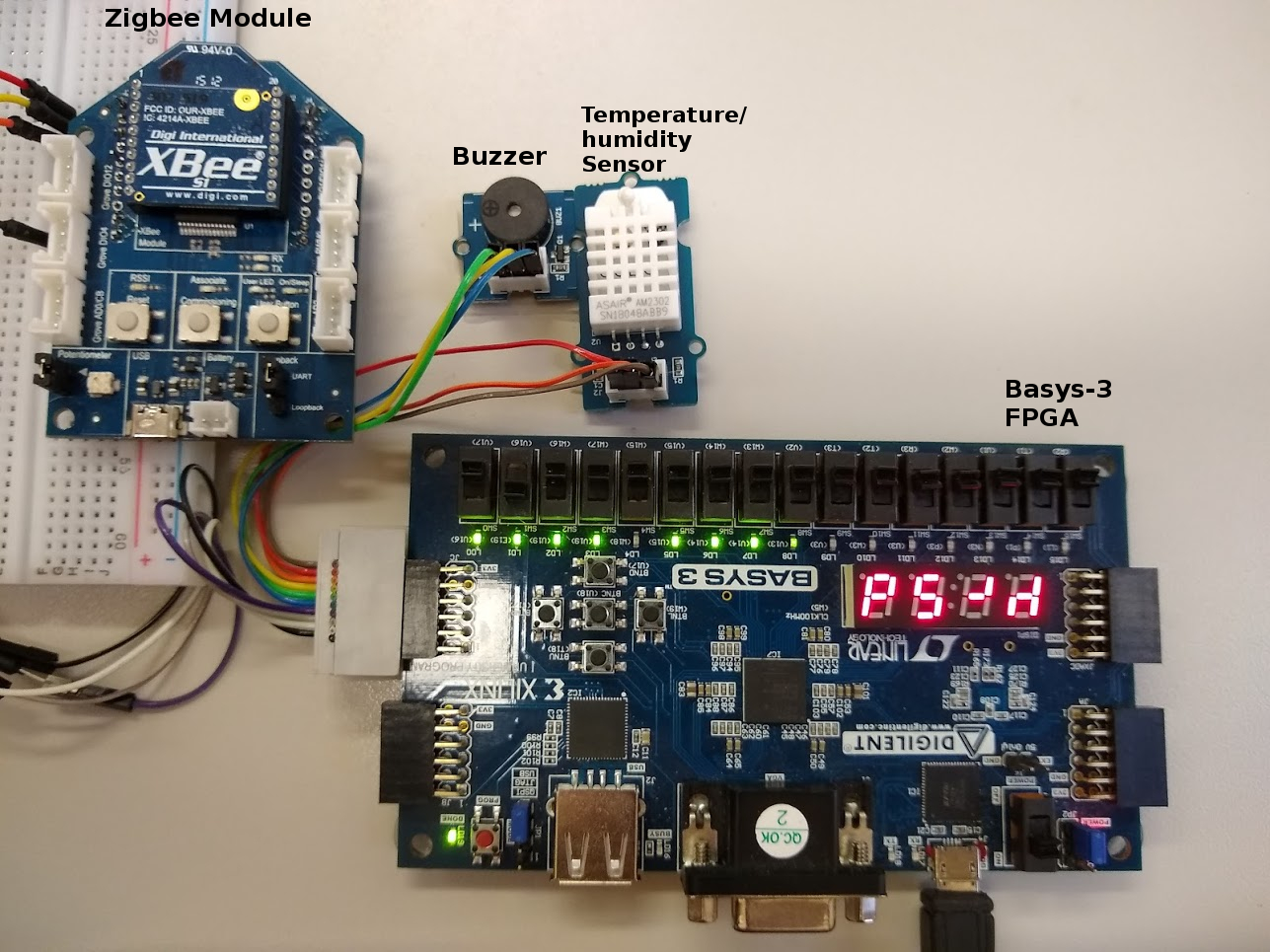}
	\caption{Hardware setup for a fire sensor}
	\label{fig:fire}
\end{figure}

\acron is used to prove execution of the fire sensor software. This software is shown in Figure~\ref{fig:sw_trans}a in 
Appendix~\ref{apdx:swt}. It consists of two main functions: \texttt{ReadSensor} and \texttt{SoundAlarm}.
Proofs of execution are requested by the safety authority via \request to issue commands to execute these functions. 
\texttt{ReadSensor} reads and processes the value generated by temperature/humidity analog device memory-mapped 
GPIO, and copies this value to $OR$. The \texttt{SoundAlarm} function turns the buzzer on for 2 seconds, i.e., it writes 
``1'' to the memory address mapped to the buzzer, busy-waits for 2 seconds, and then writes ``0'' to the same memory location.
This implementation corresponds to the one in the open-source repository~\footnote{https://github.com/Seeed-Studio/LaunchPad\_Kit}
and was ported to a \acron-enabled MCU. The porting effort was minimal: it involved around 30 additional lines of \texttt{C} code,
mainly for re-implementing sub-functions  originally implemented as shared APIs, e.g., \texttt{digitalRead/Write}.
Finally, we transformed ported code to be compatible with \acron's \pox architecture.
Details can be found in Appendix~\ref{apdx:swt}.

\section{Conclusion}\label{sec:conclusion}
This paper introduces \acron, a novel and formally verified security service targeting low-end embedded devices.
It allows a remote untrusted prover to generate unforgeable proofs of remote software execution.
We envision \acron's use in many IoT application domains, such as authenticated sensing and actuation.
Our implementation of \acron is realized on a real embedded system platform, MSP430, synthesized on an 
FPGA, and the verified implementation is publicly available.
Our evaluation shows that \acron has low overhead for both hardware footprint and time for generating proofs of execution.

\bibliographystyle{plain}
\bibliography{references}

\vspace{1em}
\appendix

\noindent\textbf{\Large APPENDIX}

\section{Sub-Module Verification}\label{apdx:sub-modules}
\acron is designed as a set of seven sub-modules. We now describe \acron's verified implementation, by focusing on two of these sub-modules and their corresponding properties.
The Verilog implementation of omitted sub-modules is available in~\cite{public-code}.
Each sub-module enforces a sub-set of the LTL specifications in Definition~\ref{def:LTL_props}.
As discussed in Section~\ref{sec:verif}, sub-modules are designed as FSMs.
In particular, we implement them as Mealy FSMs, i.e, their output changes as a function of both the current state and current input values. 
Each FSM takes as input a subset of signals shown in Figure~\ref{fig:vape} and produces only one output -- $EXEC$ -- indicating violation 
of \pox properties.

To simplify the presentation, we do not explicitly represent the value of $EXEC$ for each state transition.
Instead, we define the following implicit representation:
\begin{compactenum}
	\item $EXEC$ is 0 whenever an FSM transitions to $NotExec$ state.
	\item $EXEC$ remains 0 until a transition leaving $NotExec$ state is triggered.
	\item $EXEC$ is 1 in all other states.
	\item \textbf{Sub-modules composition:} Since all \pox properties must simultaneously hold, the value of $EXEC$ 
	produced by \acron is the conjunction (logical $AND$) of all sub-modules' individual $EXEC$ flags.
\end{compactenum}
\begin{figure}[!ht]
\begin{center}
\noindent\resizebox{\columnwidth}{!}{%
	\begin{tikzpicture}[->,>=stealth',auto,node distance=4.0cm,semithick]
		\tikzstyle{every state}=[minimum size=1.5cm]
		\tikzstyle{every node}=[font=\large]

		\node[state, fill={rgb:black,1;white,2}] 		(A)					{$NotExec$};
		\node[state]         (B) [above of=A,align=center, yshift=-1cm]	{$notER$};
		\node[state]         (C) [left  of=A,align=center]	{$fstER$};
		\node[state]         (E) [below of=A, yshift=1cm] {$midER$};
		\node[state]         (D) [right of=A]	{$lastER$};

		\path[->,every loop/.style={looseness=8}] 
			(A) edge [out=330,in=300,looseness=8] node[above right, yshift=.2cm] {\small$otherwise$} (A)
				edge [bend right=10]  node [right] [above] {\footnotesize $PC=ER_{min}\land\neg~irq$} (C)
			(B)  edge [loop above] node {\small $(PC<ER_{min}\,\lor\,PC>ER_{max})$} (C)
				edge node [above left] {\small $PC=ER_{min}\,\land\,\neg~irq$} (C)
				edge node [right] {\small$otherwise$} (A)
			(C)  edge [loop left] node [above right, left] {\small \shortstack{$PC=ER_{min}$\\$\land\,\neg~irq$}} (C)
				edge node [below left] {\small \shortstack{$(PC>ER_{min}\,\land\,PC<ER_{max})$\\$\land\,\neg~irq$}} (E)
				edge [bend right=10] node [below] {\small$otherwise$} (A)
			(E)  edge [loop below] node {\small \shortstack{$(PC>ER_{min}\,\land\,PC<ER_{max})$\\$\land\,\neg~irq$}} (C)
				edge node [below right] {\small $PC=ER_{max}\,\land\,\neg~irq$} (D)
				edge node [left] {\small$otherwise$} (A)
			(D)  edge [loop right] node [above left, right]  {\small \shortstack{$PC=ER_{max}$\\$\land\,\neg~irq$}} (D)
				edge node [above right] {\small \shortstack{$(PC<ER_{min}\,\lor\,PC>ER_{max})$\\$\land\,\neg~irq$}} (B)
				edge node [above]  {\small$otherwise$} (A);
	\end{tikzpicture}
}
\caption{Verified FSM for LTLs~\ref{eq:ephe_atom1}-\ref{eq:ephe_atom3}, a.k.a., EP2- Ephemeral Atomicity.}
\label{fig:atomicity_fsm}
\end{center}
\end{figure}
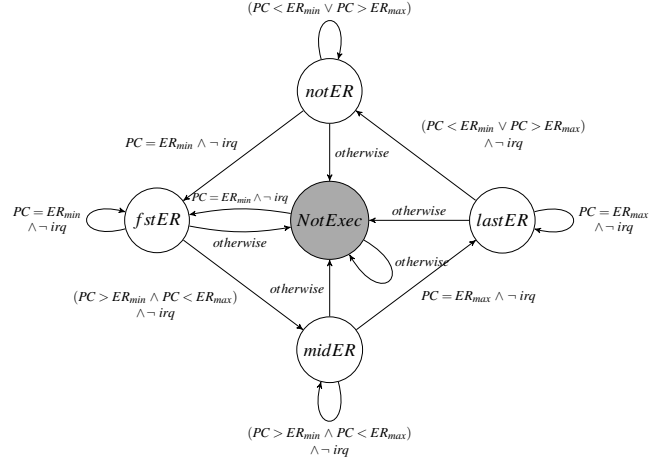

Figure~\ref{fig:atomicity_fsm} represents a verified model enforcing LTLs~\ref{eq:ephe_atom1}-\ref{eq:ephe_atom3}, 
corresponding to the high-level property \textbf{EP2- Ephemeral Atomicity}.
The FSM consists of five states.
$notER$ and $midER$ represent states when $PC$ is: (1) outside $ER$, and (2) within $ER$
respectively, excluding the first ($ER_{min}$) and last ($ER_{max}$) instructions.
Meanwhile, $fstER$ and $lstER$ correspond to states when $PC$ points to the first and last instructions, respectively. 
The only possible path from $notER$ to $midER$ is through $fstER$. Similarly, the only
path from $midER$ to $notER$ is through $lstER$. 
A transition to the $NotExec$ state is triggered whenever:
(1) any sequence of values for $PC$ do not follow the aforementioned conditions,
or (2) $irq$ is logical 1 while $PC$ is inside $ER$.
Lastly, the only way to transition out of the $NotExec$ state is 
to restart $ER$'s execution.

\begin{figure}
\begin{center}
\noindent\resizebox{0.8\columnwidth}{!}{%
	\begin{tikzpicture}[->,>=stealth',auto,node distance=8.0cm,semithick]
		\tikzstyle{every state}=[minimum size=2cm]
		\tikzstyle{every node}=[font=\large]

		\node[state] 		(A)					{$Run$};
		\node[state, fill={rgb:black,1;white,2}]         (B) [right of=A,align=center]	{$NotExec$};

		\path[->,every loop/.style={looseness=8}] 
			(A) edge [loop above] node {$otherwise$} (A)
			(B) edge [loop above] node {$otherwise$} (B);
  		
\draw[->] (A.345) -- node[rotate=0,below, align=center,auto=right] {\scriptsize \shortstack{$[W_{en} \land (D_{addr} \in METADATA)] \lor$\\$[DMA_{en} \land (DMA_{addr} \in METADATA)]$}} (B.195);
\draw[<-] (A.15) -- node[rotate=0,above] {\scriptsize \shortstack{$PC=ER_{min}\land$\\$\neg[W_{en} \land (D_{addr} \in METADATA)] \land$\\$\neg[DMA_{en} \land (DMA_{addr} \in METADATA)]$}} (B.165);
	\end{tikzpicture}
}
\caption{Verified FSM for LTL~\ref{eq:metadata_prot}, a.k.a., MP3- Challenge Temporal Consistency.}
\label{fig:meta_fsm}
\end{center}
\end{figure}
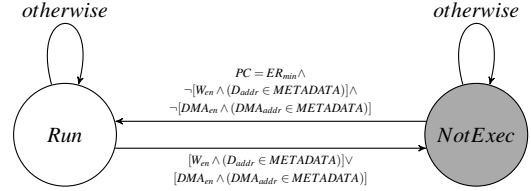

Figure~\ref{fig:meta_fsm} shows the FSM verified to comply with LTL~\ref{eq:metadata_prot} 
(\textbf{MP3- Challenge Temporal Consistency}).  The FSM has two states: $Run$ and $NotExec$. 
The FSM transitions to the $NotExec$ state and outputs $EXEC = 0$ whenever a violation happens, 
i.e., whenever $METADATA$ is modified in software.
It transitions back to $Run$ when $ER$'s execution is restarted without such violation. 
\section{Proofs for Implementation Correctness \& Security}\label{apdx:prove}
In this section we discuss the computer proof for \acron's implementation correctness (Theorem~\ref{th:vape_corr}) and the reduction proof that \acron is a secure \pox architecture as long as \vrased is a secure \RA architecture (Theorem~\ref{th:vape_sec}).
\begin{figure}[!h]
\begin{mdframed}
\begin{theorem}\label{th:vape_corr}
$\text{Definition~\ref{def:exec_model}} \land \text{LTLs~\ref{eq:ephe_immut} --\ref{eq:reset_prot}} \rightarrow \text{Definition~\ref{def:vape_fc}}$.
\end{theorem}
\end{mdframed}
\end{figure}
A formal LTL computer proof for Theorem~\ref{th:vape_corr} is available at~\cite{public-code}.
We here discuss the intuition behind such proof.
Theorem~\ref{th:vape_corr} states that LTLs~\ref{eq:ephe_immut} -- \ref{eq:reset_prot}, when considered in conjunction with the machine model in Definition~\ref{def:exec_model}, imply \acron's implementation correctness.

Recall that Definition~\ref{def:vape_fc} states that, in order to have $EXEC=1$ during the computation of \prove, at least once \textbf{before such time} the following must have happened:
\begin{enumerate}
 \item The system reached state $S_0$ in which the software stored in $ER$ started executing from its first instruction ($PC = ER_{min}$).
 \item The system eventually reached a state $S_1$ when ER finished executing ($PC = ER_{max}$). In the interval between $S_0$ and $S_1$ $PC$ remained executing instructions within $ER$, there were no interrupts, no resets, and DMA remained inactive.
 \item The system eventually reached a state $S_2$ when \prove started executing ($PC = CR_{min}$). In the interval between $S_0$ and $S_2$ the memory regions of $METADATA$ and $ER$ were not modified.
 \item In the interval between $S_0$ and $S_2$ the $OR$ memory region was only modified by $ER$'s software execution ($PC \in ER \lor \neg~\text{Modify\_Mem}(OR)$).
\end{enumerate}

The first two properties to be noted are LTL~\ref{eq:reset_prot} and LTL~\ref{eq:EXEC_ZERO}.
LTL~\ref{eq:reset_prot} establishes the default state of $EXEC$ is $0$.
LTL~\ref{eq:EXEC_ZERO} enforces that the only possible way to change $EXEC$ from $0$ to $1$ is by having $PC = ER_{min}$.
In other words, $EXEC$ is $1$ during the computation of \prove only if, at some before that, the code stored in ER started to execute (state $S_0$).

To see why state $S_1$ (when ER execution finishes, i.e., $PC = ER_{max}$) is reached and until then ER executes atomically, we look at LTLs~\ref{eq:ephe_atom1},~\ref{eq:ephe_atom2},~\ref{eq:ephe_atom3}, and~\ref{eq:no_overlap}.
LTLs~\ref{eq:ephe_atom1},~\ref{eq:ephe_atom2} and \ref{eq:ephe_atom3} enforce that $PC$ will stay inside $ER$ until $S_1$ or otherwise $EXEC$ will be set to $0$.
On the other hand, it is impossible to execute instructions of \prove ($PC \in CR$) without leaving $ER$, because LTL~\ref{eq:no_overlap} guarantees that $ER$ and $CR$ do not overlap, or $EXEC = 0$.

So far we have argued that to have a token \token that reflects $EXEC = 1$ the code contained in $ER$ must have executed successfully.
What remains to be shown is: producing this token implies the code in $ER$ and $METADATA$ are not modified in the interval between $S_0$ and $S_2$
and only $ER$'s execution can modify $OR$ in the same time interval.

Clearly, the contents of $ER$ cannot be modified after $S_0$ because Modify\_Mem($ER$) directly implies that LTL~\ref{eq:ephe_immut} will set $EXEC = 0$.
The same reasoning is applicable for modifications to $METADATA$ region with respect to LTL~\ref{eq:metadata_prot}.
The same argument applies to modifying $OR$, with the only exception that $OR$ modifications are allowed only by the CPU and when $PC \in ER$ (LTL~\ref{eq:output_prot}).
This means that $OR$ can only be modified by the execution of $ER$. In addition, LTL~\ref{eq:output_prot} also ensures that DMA is disabled during the execution of $ER$ to prevent unauthorized modification of intermediate results in data memory.
Therefore, the timeline presented in Figure~\ref{fig:memconsist} is strictly implied by \acron's implementation.
This concludes the reasoning behind Theorem~\ref{th:vape_corr}.
\begin{figure}[!h]
\begin{mdframed}
\begin{theorem}\label{th:vape_sec}
\acron is secure according to Definition~\ref{def:pox_sec_def} as long as \vrased is a secure \RA architecture according to Definition~\ref{def:vrased_sec}.
\end{theorem}
\end{mdframed}
\end{figure}
\begin{figure}[!ht]
\begin{mdframed}
\begin{definition}{\vrased's Security Game~\cite{vrasedp}}\label{def:vrased_sec}~\\
\footnotesize
\textbf{\ref{def:vrased_sec}.1 \RA Security Game (\RA-game):}

\textbf{Notation:}\\
- $l$ is the security parameter and $|\attkey| = |\chal| = |MR| = l$\\
- $AR(t)$ denotes the content of $AR$ at time $t$\\
\texttt{\RA-game:}
	\begin{enumerate}
	\item \texttt{Setup:} \adv\ is given oracle access to \sw calls.
	\item \texttt{Challenge:} A random challenge $\chal \leftarrow \$\{0,1\}^l$ is generated and given to \adv.
	\item \texttt{Response:} \adv\ responds with a pair $(M, \sigma)$, where $\sigma$ is either forged by \adv, or is 
		the result of calling \sw at some arbitrary time $t$.
	\item \adv\ wins if and only if $M \neq AR(t)$ and $\sigma = HMAC(KDF(\attkey, \chal), M)$.
	\end{enumerate}
	
\textbf{\ref{def:vrased_sec}.2 \RA Security Definition:}\\
An \RA scheme is considered secure if for all PPT adversaries \adv, there exists a negligible
function $\negl[]$ such that:
	\begin{center}
		$Pr[\adv, \text{\RA-game}] \leq \negl[l]$
	\end{center}
\end{definition}
\end{mdframed}
\end{figure}
%
\begin{proof}
\begin{small}
Assume that $\adv_{\pox}$ is an adversary capable of winning the security game in Definition~\ref{def:pox_sec_def} against \acron with more than negligible probability.
We show that, if such $\adv_{\pox}$ exists, then it can be used to construct (in a polynomial number of steps) $\adv_{\RA}$ that wins \vrased's security game (Definition~\ref{def:vrased_sec}) with more than negligible probability.
Therefore, by contradiction, nonexistence of $\adv_{\RA}$ (i.e., \vrased's security) implies nonexistence of $\adv_{\pox}$ (\acron's security).

First we recall that to win \acron's security game $\adv_{\pox}$ must provide ($\token_{\adv}$, $\xout_{\adv}$), such that $\vrfy(\token_{\adv}, \xout_{\adv}, \xsw, \chal, \cdot) = 1$.
To comply with conditions 3.a and 3.b in Definition~\ref{def:pox_sec_def}, this must be done either of the following two cases:
\begin{itemize}
	\item[\texttt{Case1}] \adv does not execute $\xsw$ in the time window between $t_{req}$ and $t_{verif}$ (i.e., $\neg \execute^{\dev}(\xsw,t_{req} \rightarrow t_{verif})$).
	\item[\texttt{Case2}] \adv calls $\execute^{\dev}(\xsw,t_{req} \rightarrow t_{verif})$ but modifies its output \xout in between the time when the execution of \xsw completes and the time when \prove is called.
\end{itemize}

However, according to the specification of \acron's $\vrfy$ algorithm (see Definition~\ref{def:vape})
a token $\token_{\adv}$ will only be accepted if it reflects an input value with $EXEC = 1$, as expected by \vrf.
In \acron's implementation \xout is stored in region $OR$, and \xsw in region $ER$.
Moreover, given Theorem~\ref{th:vape_corr}, we know that having $EXEC = 1$ during \prove implies three conditions have been fulfilled:
\begin{enumerate}
	\item[\texttt{Cond1}] The code in $ER$ executed successfully.
	\item[\texttt{Cond2}] The code in $ER$ and $METADATA$ were not modified after starting $ER$'s execution and before calling \prove.
	\item[\texttt{Cond3}] Outputs in $OR$ were not modified after completing $ER$'s execution and before calling \prove.
\end{enumerate}
The third condition rules out the possibility of \texttt{Case2} since that case assumes \adv can modify \xout, resided in $OR$, after $ER$ execution and $EXEC$ stays logical 1 during \prove. 
We further break down \texttt{Case1} into three sub-cases:

\begin{itemize}
	\item[\texttt{Case1.1}] \adv does not follow Cond1-Cond3. The only way for \adv to produces ($\token_{\adv}$, $\xout_{\adv}$) in this case is to not call $\prove$, i.e., by directly guessing \token.
	\item[\texttt{Case1.2}] \adv follows Cond1-Cond3 but does not execute \xsw between $t_{req}$ and $t_{verif}$. Instead, it produces  ($\token_{\adv}$, $\xout_{\adv}$) 
	by calling:
\begin{equation}
\xout_{\adv} \equiv \execute^{\dev}(ER_{\adv},t_{req} \rightarrow t_{verif})
\end{equation}

where $ER_{\adv}$ is a memory region different from the one specified by \vrf on \request ($\adv_{\pox}$ can do this by modifying $METADATA$ to different values of $ER_{min}$ and $ER_{max}$ before calling \execute). 
	\item[\texttt{Case1.3}] Similar to Case1.2, but $ER_{\adv}$ is the same region specified by \vrf on \request containing a different executable $\xsw_{\adv}$.
\end{itemize}

We show that an adversary that succeeds in any of these cases can be used win \vrased's security game.
To see why this is the case, we note that \acron's \prove function is implemented by using \vrased's \sw without any modification.
\sw covers memory regions $MR$ (challenge memory) and $AR$ (attested region).
Hence, \acron instantiates these memory regions as:
\begin{enumerate}
 \item $MR = \chal$;
 \item $ER \subset AR$;
 \item $OR \subset AR$;
 \item $METADATA \subset AR$;
\end{enumerate}

Doing so ensures that all sensitive memory regions used by \acron are included among the inputs to \vrased's attestation. 
Let $X(t)$ denote the content in memory region $X$ at time $t$.
$\adv_{\RA}$ can then be constructed using $\adv_{\pox}$ as follows:
\begin{enumerate}
 \item $\adv_{\RA}$ receives \chal from the challenger in step (2) of \RA  security game of Definition~\ref{def:vrased_sec}.
 \item At arbitrary time $t$, $\adv_{\RA}$ has 3 options to write $AR(t) = AR_{\adv}$ and call $\adv_{\pox}$:
 
 \begin{enumerate}
  \item Modify $ER(t) \neq \xsw$ or $OR(t) \neq \xout$ or $METADATA(t) \neq METADATA_{\vrf}$. It then calls $\adv_{\pox}$ in \texttt{Case1.1}.
  \item Modify $ER$ to be different from the range chosen by \vrf. Therefore, $METADATA(t) \neq METADATA_{\vrf}$. It then calls $\adv_{\pox}$ in \texttt{Case1.2}.
  \item Modify $ER(t)$ to be different from \xsw. It then calls $\adv_{\pox}$ in \texttt{Case1.3}.
 \end{enumerate}
 

 In any of these options, $\adv_{\RA}$ will produce ($\token_{\adv}$,$\xout_{\adv}$), such that $\vrfy(\token_{\adv}, \xout_{\adv}, \xsw, \chal, \cdot) = 1$ with non-negligible probability.
 \item $\adv_{\RA}$ replies to the challenger with the pair $(M, \token_{\adv})$, where $M$ corresponds to the values of $\xsw$, $\xout$ and $METADATA_{\vrf}$, matching $\token_{\adv}$ and $\xout_{\adv}$ generated by $\adv_{\pox}$.
 By construction $M \neq AR_{\adv}= AR(t)$, as required by Definition~\ref{def:vrased_sec}.
 \item Challenger will accept $(M, \token_{\adv})$ with the same non-negligible probability that $\adv_{\pox}$ has of producing ($\token_{\adv}$,$\xout_{\adv}$) such that $\vrfy(\token_{\adv}, \xout_{\adv}, \xsw, \chal, \cdot) = 1$.
\end{enumerate}
\end{small}
\end{proof}
%
%
\lstset{language=C,
	basicstyle={\scriptsize\ttfamily},
	showstringspaces=false,
	frame=single,
	xleftmargin=2em,
	framexleftmargin=3em,
	numbers=left, 
	numberstyle=\tiny,
	commentstyle={\tiny\itshape},
	keywordstyle={\tiny\bfseries},
	keywordstyle=\color{blue}\tiny\ttfamily,
	stringstyle=\color{red}\tiny\ttfamily,
        commentstyle=\color{black}\tiny\ttfamily,
        morecomment=[l][\color{magenta}]{\#},
        breaklines=true
}
\begin{figure*}
\centering
\begin{minipage}{.8\linewidth}
\begin{lstlisting}[basicstyle=\tiny, numberstyle=\tiny]
#define  P4IN        (*(volatile unsigned char *) 0x001C)
#define  P4OUT       (*(volatile unsigned char *) 0x001D)
#define  P4DIR       (*(volatile unsigned char *) 0x001E)
#define  P4SEL       (*(volatile unsigned char *) 0x001F)
#define  BIT4        (0x0010)

#define MAXTIMINGS   85

#define OR           0xEEE0 // OR is in AR

#define HIGH         0x1
#define LOW          0x0
#define INPUT        0x0
#define OUTPUT       0x1

__attribute__ ((section (".exec.entry"), naked)) void ReadSensorEntry() {
    // ERmin
    ReadSensor();
    __asm__ volatile( "br #__exec_leave" "\n\t");
}

__attribute__ ((section (".exec.body"))) int digitalRead() {
    if(P3IN & BIT4) return HIGH;
    else return LOW;
}

__attribute__ ((section (".exec.body"))) void digitalWrite(uint8_t val) {
    if (val == LOW)
        P3OUT &= ~BIT4;
    else
        P3OUT |= BIT4;
}

__attribute__ ((section (".exec.body"))) void pinMode(uint8_t val) {
    if (val ==  INPUT)
        P3DIR &= ~BIT4;
    else if (val == OUTPUT)
        P3DIR |= BIT4;
}

__attribute__ ((section (".exec.body"))) void ReadSensor() {
    // Tell the sensor that we are about to read
    digitalWrite(HIGH);
    delayMS(250);
    pinMode(OUTPUT);
    digitalWrite(LOW);
    delayMS(20);
    digitalWrite(HIGH);
    delayMicroseconds(40);
    pinMode(INPUT);
    uint8_t laststate = HIGH, counter = 0, j = 0, i;
    uint8_t data[5] = {0};
    // Read the sensor's value
    for ( i=0; i< MAXTIMINGS; i++) {
        counter = 0;
        while (digitalRead() == laststate) {
          counter++;
          if (counter == 255) {
            break;
          }
        }
        laststate = digitalRead();
        if (counter == 255) break;
        if ((i >= 4) && (i%2 == 0)) {
          data[j/8] <<= 1;
          if (counter > 100) {
            data[j/8] |= 1;
            avg += counter;
            k++;
          }
          j++;
        }

    }
    // Copy the reading to OR
    memcpy(OR, data, 5);
}

__attribute__ ((section (".exec.exit"), naked)) void ReadSensorExit() {
    __asm__ volatile("ret" "\n\t");
    // ERmax
}

\end{lstlisting}
\centering
(a) Fire Sensor's code written in \texttt{C}
\end{minipage}
\hfill
\begin{minipage}{\linewidth}
\begin{lstlisting}[basicstyle=\tiny, numberstyle=\tiny, xleftmargin=.13\textwidth, xrightmargin=.13\textwidth]
...
SECTIONS
{
  ...
  .text :
  {
     ...
    *(.exec.entry)
    . = ALIGN(2);
    *(.exec.body)
    . = ALIGN(2);
    PROVIDE (__exec_leave = .);
    *(.exec.exit)
  }  > REGION_TEXT
  ...
}
...
\end{lstlisting}
\centering
(b) Linker script
\end{minipage}
\caption{Code snippets for (a) fire sensor described in Section~\ref{sec:auth_sensing} (b) linker script}
\label{fig:sw_trans}%
\end{figure*}


\section{Executable Limitations}\label{apdx:limit}

We now discuss the limitations of our approach on the executable types.

\noindent\textbf{Shared libraries.}
In order to produce a valid proof, \vrf must ensure that execution of \xsw
does not depend on external code located outside its execution range $ER$ (e.g., shared libraries).
A call to such code would violate LTL~\ref{eq:ephe_atom1}, 
resulting in $EXEC = 0$ during the HMAC computation. 
One possible way to support this type of executable is to transform it into a self-contained executable by statically linking all dependencies 
during the compilation time. Another is to appropriately set $ER$ to cover all external code used by \xsw.


\noindent\textbf{Self-modifying code (SMC).} SMC is a type of executable that alters itself while executing.
Clearly, this executable type violates LTL~\ref{eq:ephe_immut} 
that requires the code in $ER$ to remain unchanged during
$ER$ execution. It is unclear how \acron can be adapted to support SMC; however, we are unaware of any legitimate and realistic use-case of SMC 
in our target bare-metal applications.

\noindent\textbf{Interrupts.} 
Our notion of successful execution in Section~\ref{sec:arch} prohibits an interrupt to happen during \xsw's execution.
This limitation can be problematic especially for interrupt-driven programs such as the ones in real-time systems. 
Nonetheless, simply allowing interrupts to happen during the execution may result in attacks that allow
malware to modify intermediate execution results in data memory and consequently influence the execution output. 
One possible way to remedy this issue is to allow interrupts as long as 
all interrupt handlers are: (1) immutable from the start of execution till the end of attestation and
(2) included in the attested memory range during the attestation process.
\vrf then can determine whether an interrupt that may have happened during the execution is malicious
by inspecting all interrupt handlers from the proof of execution.


\section{Software Transformation}\label{apdx:swt}

Recall that our notion of successful execution (in Section~\ref{sec:arch}) requires the function's entry point to be at the first instruction and the exit point to be at the last instruction.
In this section, we discuss an efficient way to transform arbitrary software (besides the ones in Appendix~\ref{apdx:limit}) implementing a function to conform with this requirement.

Line 10-17 of Figure~\ref{fig:sw_trans} shows an (partial) implementation of the \texttt{ReadSensor} function described in Section~\ref{sec:auth_sensing}.
This implementation, when converted to an executable, does not guarantee \acron's executable requirement since the compiler
may choose to place one of its sub-functions, instead of \texttt{ReadSensor}, to the entry and/or exit points of the executable.
One obvious way to fix this issue is to implement all of its sub-functions as inline functions;
however, such approach may be inefficient as in this example it will create multiple duplicate code for the same sub-functions (e.g., \texttt{digitalWrite}) inside the executable.

Instead, we created the dedicated functions for the entry (Line 1-4) and exit (Line 6-8) points, and assign those functions to separated executable sections -- 
``.exec.entry'' for the entry and ``.exec.exit'' for the exit.
Then, we labeled all sub-functions used by \texttt{ReadSensor} as well as \texttt{ReadSensor} itself to the same section -- ``.exec.body'' -- and 
modified the MSP430 linker to place ``.exec.body'' between ``.exec.entry'' and ``.exec.exit'' sections. The modified linker script is shown in Figure~\ref{fig:sw_trans}b.
This way, we ensure that the entry and exit function locate at the beginning and the end of the executable, respectively, and thus the resulting executable conforms with \acron's requirement.



\end{document}